\documentclass[11pt, draftcls, letterpaper, onecolumn]{IEEEtran}

\usepackage{amsmath}
\usepackage{amssymb}
\usepackage{graphicx}
\usepackage{cite}
\usepackage{mathrsfs}
\usepackage{stmaryrd}
\usepackage{hyperref}
\usepackage{subfigure}

\newtheorem{theorem}{Theorem}
\newtheorem{corollary}{Corollary}[theorem]
\newtheorem{lemma}{Lemma}
\newtheorem{proposition}{Proposition}
\newtheorem{definition}{Definition}

\newtheorem{remark}{Remark}
\newtheorem{example}{Example}

\newcommand{\mbf}[1]{\mathbf{#1}}
\newcommand{\set}[1]{\mathscr{#1}}
\newcommand{\reals}{\mathbb{R}_+}

\graphicspath{{.}{./Figures/}}

\title{Lossy Broadcasting in Two-Way Relay Networks with Common Reconstructions}
\author{Roy Timo, Alex Grant and Gerhard Kramer
\thanks{The work of R. Timo and A. Grant was supported by the Australian Research Council Grant DP0880223.}
\thanks{R. Timo and A. Grant are with the Institute for Telecommunications Research at the University of South Australia. Email: \{roy.timo, alex.grant\}@unisa.edu.au.}
\thanks{G. Kramer is with the Department of Electrical Engineering and Information Technology at the Technische Universit\"{a}t M\"{u}nchen. Email: gerhard.kramer@tum.de.}
}

\begin{document}

\maketitle

\begin{abstract}
The broadcast phase (downlink transmission) of the two-way relay network is studied in the source coding and joint source-channel coding settings. The rates needed for reliable communication are characterised for a number of special cases including: small distortions, deterministic distortion measures, and jointly Gaussian sources with quadratic distortion measures. The broadcast problem is also studied with common-reconstruction decoding constraints, and the rates needed for reliable communication are characterised for all discrete memoryless sources and per-letter distortion measures.
\end{abstract}


\newpage


\section{Introduction}\label{Sec:1}

\begin{figure}[h]
\centering
\subfigure[Phase $1$ (uplink)]{
\includegraphics[width=55mm]{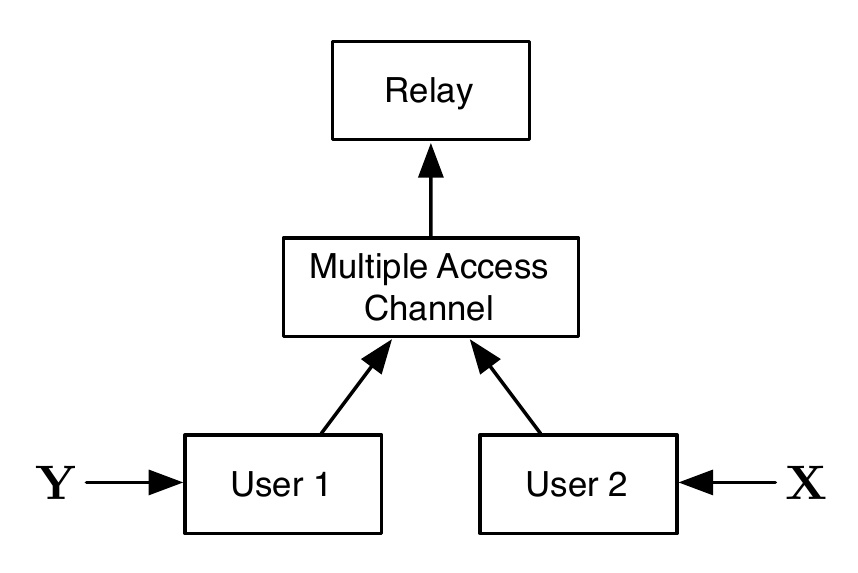}
\label{Fig:Relay-1a}
}
\subfigure[Phase $2$ (downlink)]{
\includegraphics[width=55mm]{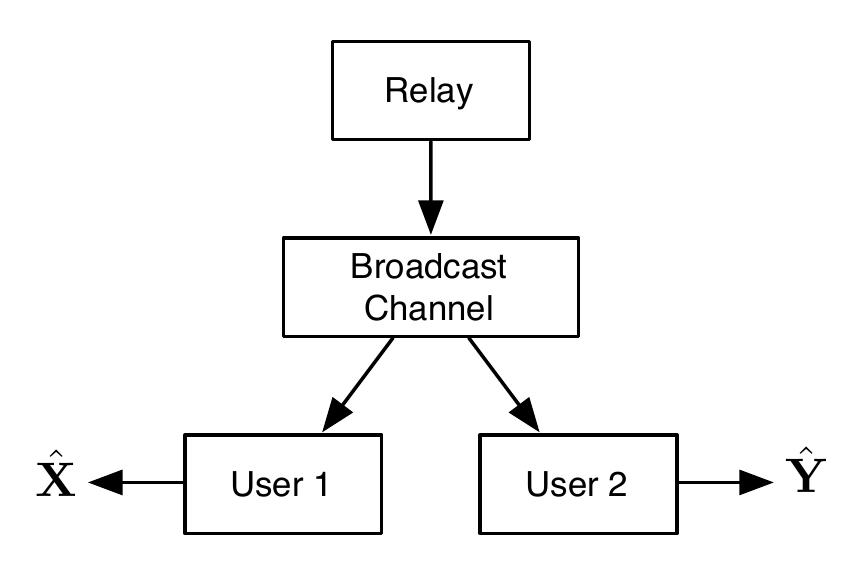}
\label{Fig:Relay-1b}
}
\caption{The two-way relay network: user $1$ has $\mbf{Y}$ and requires a copy $\hat{\mbf{X}}$ of $\mbf{X}$ from user $2$; similarly, user $2$ has $\mbf{X}$ and requires a copy $\hat{\mbf{Y}}$ of $\mbf{Y}$ from user $1$. Figure~\ref{Fig:Relay-1a} depicts the uplink and Figure~\ref{Fig:Relay-1b} depicts the downlink.}
\label{Fig:Relay}
\end{figure}

\IEEEPARstart{C}{onsider} the two-way relay network shown in Figure~\ref{Fig:Relay}. User $1$ requires an approximate copy $\hat{\mbf{X}}$ of the data $\mbf{X}$ from user $2$, and user $2$ requires an approximate copy $\hat{\mbf{Y}}$ of the data $\mbf{Y}$ from user $1$. The users are physically separated and direct communication is not possible. Instead, indirect communication is achieved via a relay and a two-phase communication protocol. In phase $1$ (uplink transmission), each user encodes its data to a codeword that is transmitted over a multiple access channel to the relay. In phase $2$ (downlink transmission), the relay completely or partly decodes the noise-corrupted codewords it receives from the multiple access channel, and it transmits a new codeword over a broadcast channel to both users. From this broadcast transmission, user $1$ decodes $\hat{\mbf{X}}$ and user $2$ decodes $\hat{\mbf{Y}}$.

\begin{figure}[h]
\centering
\subfigure[]{
\includegraphics[width=40mm]{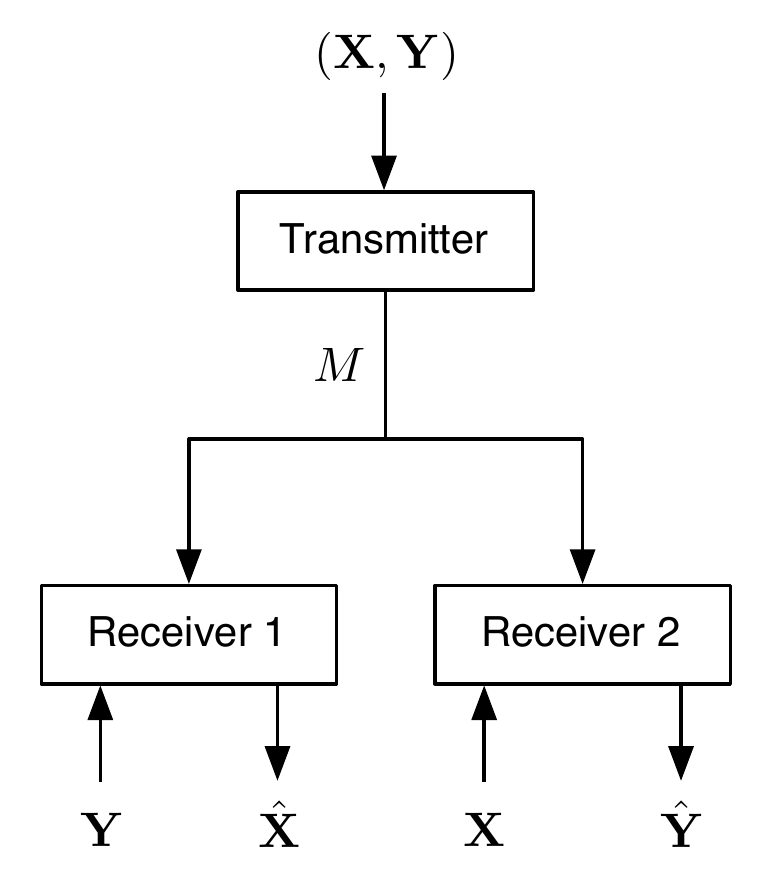}
\label{Fig:Lossy-Broadcast-1a}
}
\subfigure[]{
\includegraphics[width=40mm]{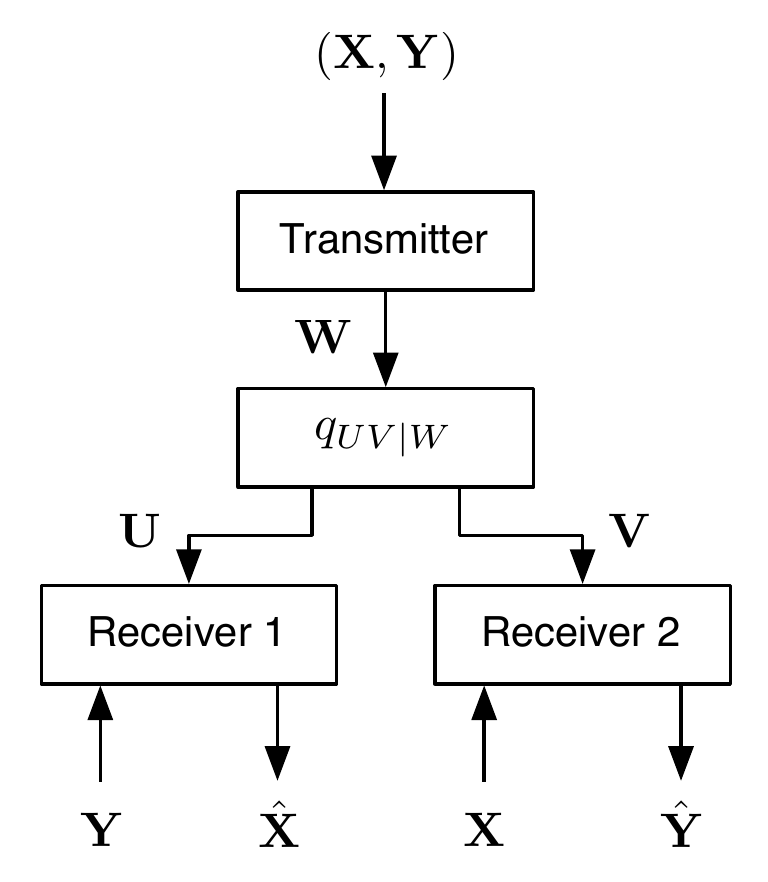}
\label{Fig:Lossy-Broadcast-1b}
}
\caption{Lossy broadcasting in two-way relay networks. The source coding and joint source-channel coding problems are shown in Figures~\ref{Fig:Relay-1a} and~\ref{Fig:Relay-1b}, respectively.}
\label{Fig:Lossy-Broadcast}
\end{figure}

In this paper, we study the downlink for the case where $\mbf{X}$ and $\mbf{Y}$ have been perfectly decoded by the relay after the uplink transmission (Figure~\ref{Fig:Lossy-Broadcast}). We are interested in the lossy setting where $\hat{\mbf{X}}$ and $\hat{\mbf{Y}}$ need to satisfy average distortion constraints. We have a source coding problem  (Figure~\ref{Fig:Lossy-Broadcast-1a}) when the broadcast channel is noiseless, and we have a joint source-channel coding problem when the broadcast channel is noisy (Figure~\ref{Fig:Lossy-Broadcast-1b}). In Figure~\ref{Fig:Lossy-Broadcast} we have relabelled the relay as the transmitter, user $1$ as receiver $1$ and user $2$ as receiver $2$. We note that the source coding problem is a special case of the joint source-channel coding problem; however, we will present each problem separately for clarity.

It is worthwhile to briefly discuss some of the implicit assumptions in the two-way relay network setup. The no direct communication assumption has been adopted by many authors including Oechtering, \textit{et al.}~\cite{Oechtering-Jan-2008-A,Oechtering-Sep-2008-A}, G\"{u}nd\"{u}z, Tuncel and Nayak~\cite{Gunduz-Sep-2008-C} as well as Wyner, Wolf and Willems~\cite{Wyner-Jun-2002-A}. It is appropriate when the users are separated by a vast physical distance and communication is via a satellite. It is also appropriate when direct communication is prevented by practical system considerations. In cellular networks, for example, two mobile phones located within the same cell will communicate with each other via their local base-station. We note that this assumption differs from Shannon's classic formulation of the two-way communication problem~\cite{Shannon-1961-C,van-der-Meulen-Jan-1977-A}. Specifically, those works assume that the users exchange data directly over a discrete memoryless channel without using a relay. The two-phase communication protocol assumption (uplink and downlink) is appropriate when the users and relay cannot transmit and receive at the same time on the same channel~\cite{Oechtering-Jan-2008-A,Wu-Mar-2005-C}. This again contrasts to Shannon's two-way communication problem~\cite{Shannon-1961-C} as well as G\"{u}nd\"{u}z, Tuncel and Nayak's separated relay~\cite{Gunduz-Sep-2008-C}, where simultaneous transmission and reception is permitted. Finally, this relay network is restricted in the sense that it does not permit feedback~\cite{Shannon-1961-C}; that is, each user cannot use previously decoded data when encoding new data.

\textit{Notation:} The non-negative real numbers are written $\reals$. Random variables and random vectors are identified by uppercase and bolded uppercase letters, respectively. The alphabet of a random variable is identified by matching calligraphic typeface, and a generic element of an alphabet is identified by a matching lowercase letter. For example, $X$ represent a random variable that takes values $x$ from a finite alphabet $\set{X}$, and $\mbf{X} = X_1,X_2,\ldots,X_n$ denotes a vector of random variables with each taking values from $\set{X}$. The length of a random vector will be clear from context. The $n$-fold Cartesian product of a single set is identified by a superscript $n$. For example, $\set{X}^n$ is the $n$-fold product of $\set{X}$. 

\textit{Paper Outline:} In Section~\ref{Sec:2}, we formally state the problem and review some basic RD functions. 
We present our main results in Section~\ref{Sec:3}, and we prove these results in Sections~\ref{Sec:4} and~\ref{Sec:5}. The paper is concluded in Section~\ref{Sec:6}.


\section{Formal Problem Statement $\&$ Definitions}\label{Sec:2}

Let $\set{X}$, $\hat{\set{X}}$, $\set{Y}$ and $\hat{\set{Y}}$ be finite alphabets, and let $q_{XY}(x,y) = \Pr[X=x,Y=y]$  be a generic probability mass function (pmf) on $\set{X} \times \set{Y}$. The source coding and joint source-channel coding problems are defined next. 

\subsection{Source Coding}

\begin{figure}[h]
\centering
\subfigure[]{
\includegraphics[width=55mm]{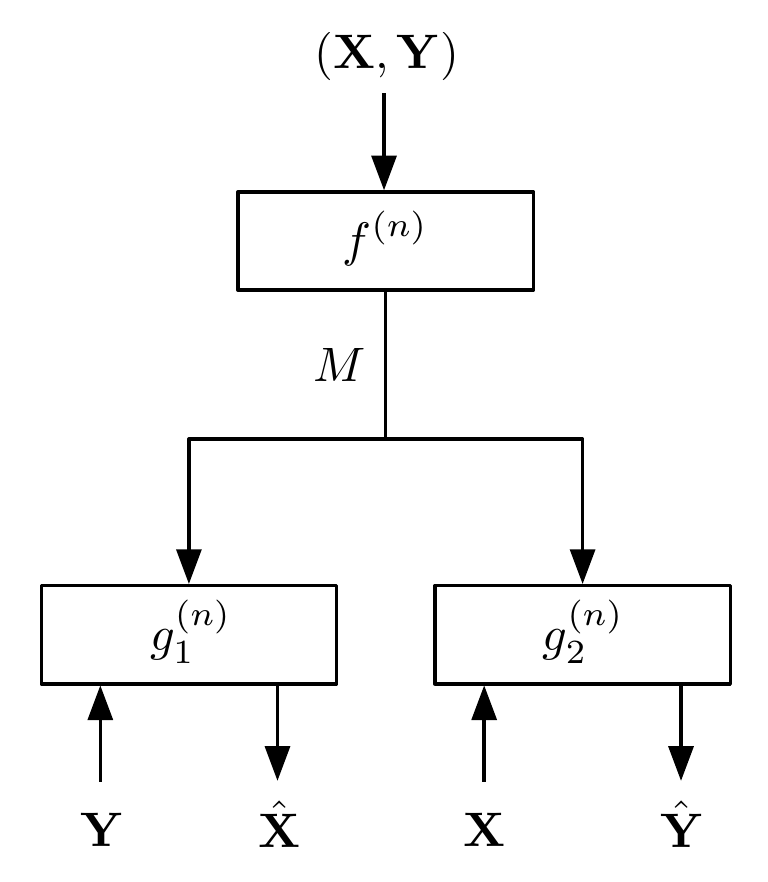}
\label{Fig:SC-Code}
}
\subfigure[]{
\includegraphics[width=52mm]{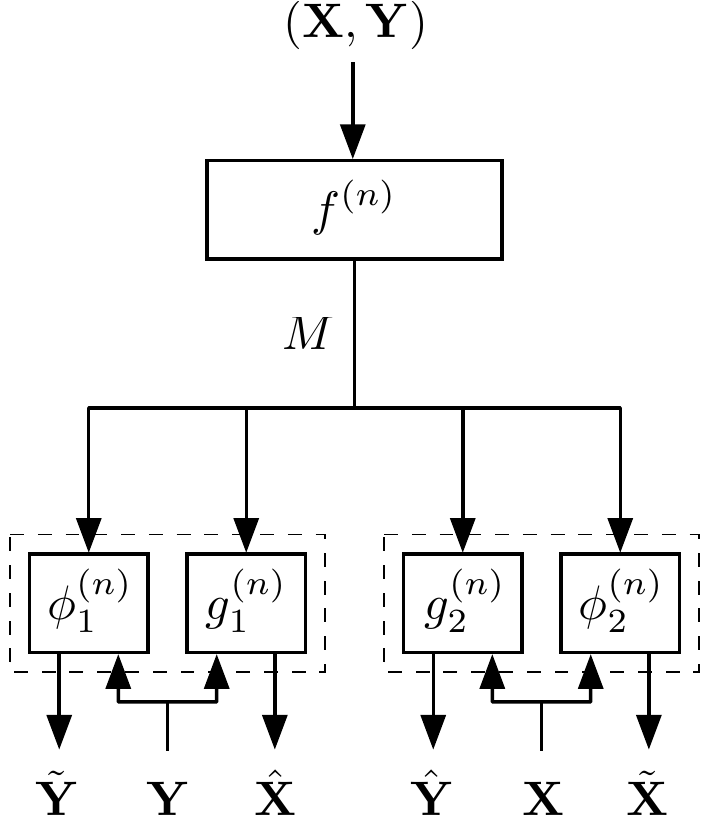}
\label{Fig:CR-SC-Code}
}
\caption{Figure (a): Encoder and decoder structure for source coding at rate $R(d_1,d_2)$. Figure (b): Encoder and decoder structure for source coding with common reconstructions at rate $R_{CR}(d_1,d_2)$.}
\end{figure}

Assume that $(\mbf{X},\mbf{Y}) = (X_1,Y_1),(X_2,Y_2), \ldots, (X_n,Y_n)$ is drawn independent and identically distributed (iid) according to $q_{XY}(x,y)$. A rate-distortion (RD) blockcode is a triple of mappings $(f^{(n)}$, $g_1^{(n)}$, $g_2^{(n)})$, where
\begin{subequations}\label{Eqn:Enc-Dec}
\begin{align}
f^{(n)} :&\ \set{X}^n \times \set{Y}^n \rightarrow \set{M}^{(n)}\ , \\
g_1^{(n)} :&\ \set{M}^{(n)} \times \set{Y}^n \rightarrow \hat{\set{X}}^n \text{ and}\\
g_2^{(n)} :&\ \set{M}^{(n)} \times \set{X}^n \rightarrow \hat{\set{Y}}^n\ .
\end{align}
\end{subequations}
Here $f^{(n)}$ denotes the encoder at the transmitter and $g_i^{(n)}$ denotes the decoder at receiver $i = 1,2$, see Figure~\ref{Fig:SC-Code}. The compression rate $\kappa^{(n)}$ of an RD code $(f^{(n)}$, $g_1^{(n)}$, $g_2^{(n)})$ is defined by
\begin{align}\label{Eqn:SC-Rate}
\kappa^{(n)} &\triangleq \frac{1}{n} \log_2 \big|\set{M}^{(n)}\big|\ ,
\end{align}
where $|\set{M}^{(n)}|$ denotes the cardinality of $\set{M}^{(n)}$. We use the braced superscript $(n)$ to emphasize that a blockcode of length $n$ is under consideration. 

The reconstruction quality of the decoded data is quantified in the usual way via average per-letter distortions. To this end, we let
\begin{subequations}
\begin{align}
\delta_1&:\ \set{X} \times \hat{\set{X}} \rightarrow [0,d_{1,\text{max}}]\quad \text{ and}\\
\delta_2&:\ \set{Y} \hspace{.8mm} \times  \hat{\set{Y}}\hspace{.8mm} \rightarrow [0,d_{2,\text{max}}]
\end{align}
\end{subequations}
be bounded per-letter distortion measures. To simplify our presentation, we assume that $\delta_1$ and $\delta_2$ are normal~\cite{Yeung-2002-B}. That is, for all $x \in \set{X}$ we have $\delta_1(x,\hat{x}) = 0$ for some $\hat{x} \in \hat{\set{X}}$. Similarly, for all $y \in \set{Y}$ we have $\delta_2(y,\hat{y}) = 0$ for some $\hat{y} \in \hat{\set{Y}}$. This assumption is not too restrictive, and our results can be extended to more general distortion measures~\cite{Yeung-2002-B}. We call $\delta_1$ a Hamming distortion measure if $\hat{\set{X}} = \set{X}$, $\delta_1(x,\hat{x}) = 0$ for $x = \hat{x}$ and $\delta_1(x,\hat{x}) = 1$ for $x \neq \hat{x}$. We call $\delta_1$ a difference distortion measure~\cite{Zamir-Nov-1996-A} if it can be written in the form $\delta_1(x-\hat{x})$, where $\hat{\set{X}} = \set{X} = \{0,1,\ldots,l_x-1\}$ and the subtraction is performed modulo-$l_x$. The same naming convention applies to $\delta_2$.

The average average distortions $(\Delta_1^{(n)},\Delta_2^{(n)})$  of an RD code $(f^{(n)}$, $g_1^{(n)}$, $g_2^{(n)})$ are defined by
\begin{subequations}\label{Eqn:SC-Dist}
\begin{align}
\label{Eqn:SC-Dist-1}
\Delta_1^{(n)} &\triangleq \mathbb{E}\left[\frac{1}{n}\sum_{i=1}^n \delta_1(X_i,\hat{X}_i) \right]\\
\label{Eqn:SC-Dist-2}
\Delta_2^{(n)} &\triangleq \mathbb{E}\left[\frac{1}{n}\sum_{i=1}^n \delta_2(Y_i\hspace{.8mm},\hat{Y}_i\hspace{.8mm}) \right],
\end{align}
\end{subequations}
where $\hat{\mbf{X}} \triangleq g^{(n)}_1(M,\mbf{Y})$, $\hat{\mbf{Y}} \triangleq g^{(n)}_2(M,\mbf{X})$, $M \triangleq f^{(n)}(\mbf{X},\mbf{Y})$, and $\mathbb{E}[\cdot]$ denotes the expectation operator.

\medskip

\begin{definition}[Source Coding]\label{Def:Source-Coding}
Let $(d_1,d_2) \in \mathbb{R}_+^2$. A rate $r \in \mathbb{R}_+$ is said to be $(d_1,d_2)$-achievable if for arbitrary $\epsilon > 0$ there exists an RD code $(f^{(n)}$, $g_1^{(n)}$, $g_2^{(n)})$ for some sufficiently large $n$ with 
\begin{subequations}
\label{Eqn:RD-Def-Ach}
\begin{align}
\kappa^{(n)}    &\leq r+\epsilon\ , \quad\text{and}\\
\Delta_i^{(n)}  &\leq d_i+\epsilon\ ,\quad i = 1,2\ .
\end{align}
\end{subequations}
Let $\set{R}(d_1,d_2)$ denote the set of all $(d_1,d_2)$-admissible rates, and let
\begin{equation}
R(d_1,d_2) \triangleq \min_{r \in \set{R}(d_1,d_2)} r\ .
\end{equation}
\end{definition}

\medskip

Definition~\ref{Def:Source-Coding} does not require that the two receivers agree on the exact realizations of $\hat{\mbf{X}}$ and $\hat{\mbf{Y}}$. For example, receiver $1$ need not know the exact realization of $\hat{\mbf{Y}}$. In some scenarios\footnote{Examples of such problems can be found in Steinberg's work~\cite{Steinberg-Nov-2009-A} on common reconstructions for the Wyner-Ziv problem.}, it is appropriate that the receivers exactly agree on $\hat{\mbf{X}}$ and $\hat{\mbf{Y}}$. The notion of common reconstructions is useful for such scenarios. 

A common-reconstructions rate-distortion (CR-RD) code is a tuple of mappings $(f^{(n)}$, $g_1^{(n)}$, $g_2^{(n)}$, $\phi_1^{(n)}$, $\phi_2^{(n)})$, where $f^{(n)}$ and $g_i^{(n)}$ are given by~\eqref{Eqn:Enc-Dec} and
\begin{subequations}\label{Eqn:CR-Enc-Dec}
\begin{align}
\phi_1^{(n)} : &\ \set{M} \times \set{Y}^n\hspace{0.8mm} \rightarrow \hspace{0.8mm} \hat{\set{Y}}^n\quad \text{ and}\\
\phi_2^{(n)} : &\ \set{M} \times \set{X}^n \rightarrow \hat{\set{X}}^n\ .
\end{align}
\end{subequations}
Here $\phi_i$ denotes the ``common-reconstruction'' decoder at receiver $i = 1,2$, see Figure~\ref{Fig:CR-SC-Code}.

The rate $\kappa^{(n)}$ and average distortion $(\Delta^{(n)}_1,\Delta^{(n)}_2)$ of a CR-RD code are defined in the same manner as~\eqref{Eqn:SC-Rate} and~\eqref{Eqn:SC-Dist}. Additionally, we define the average probability of common-reconstruction decoding error by
\begin{equation}\label{Eqn:CR-Error-Prob}
P_e \triangleq \max \big\{\Pr[\tilde{\mbf{X}} \neq \hat{\mbf{X}}],\Pr[\tilde{\mbf{Y}} \neq \hat{\mbf{Y}}]\big\}\ ,
\end{equation}
where $\tilde{\mbf{Y}} \triangleq \phi_1^{(n)}(M,\mbf{Y})$ and $\tilde{\mbf{X}} \triangleq \phi_2^{(n)}(M,\mbf{X})$.

\medskip

\begin{definition}[Source Coding with Common Reconstructions]\label{Def:CR-Source-Coding}
Let $(d_1,d_2) \in \reals^2$. A rate $r \in \mathbb{R}_+$ is said to be $(d_1,d_2)$-achievable with common reconstructions if for arbitrary $\epsilon > 0$ there exists a CR-RD code $(f^{(n)}$, $g_1^{(n)}$, $g_2^{(n)}$, $\phi_1^{(n)}$, $\phi_2^{(n)})$ with $(\kappa^{(n)}$, $\Delta_1^{(n)},$ $\Delta_2^{(n)})$ satisfying~\eqref{Eqn:RD-Def-Ach} and $P_e \leq \epsilon$. Let $\set{R}_{CR}(d_1,d_2)$ denote the set of all $(d_1,d_2)$-admissible rates with common reconstructions, and let
\begin{equation}
R_{CR}(d_1,d_2) \triangleq \min_{r \in \set{R}_{CR}(d_1,d_2)} r\ .
\end{equation}
\end{definition}

\medskip

The next proposition follows directly from Definitions~\ref{Def:Source-Coding} and~\ref{Def:CR-Source-Coding}.

\medskip

\begin{proposition}\label{Pro:RC-CR-Convex}
The RD function $R(d_1,d_2)$ and the CR-RD function $R_{CR}(d_1,d_2)$ are continuous, non-increasing and convex on $\mathbb{R}_+^2$. Moreover, for $(d_1,d_2) \in \reals^2$ we have that
\begin{equation}
R(d_1,d_2) \leq R_{CR}(d_1,d_2) \ .
\end{equation}
\end{proposition}

\textit{General Remark: } The common-reconstruction condition used in this paper was inspired by Steinberg's study~\cite{Steinberg-Nov-2009-A} of common reconstructions for the Wyner-Ziv problem. 

\subsection{Joint Source-Channel Coding}

Consider the joint source-channel coding problem. Suppose that the source $q_{XY}$ emits symbols at the rate $\kappa_s$, and that the channel accepts and emits symbols at the rate $\kappa_c$. Let $\set{W}$ denote the channel input alphabet, let $\set{U} \times \set{V}$ denote the product of the channel output alphabets, and let the transitions from $\set{W}$ to $\set{U} \times \set{V}$ be governed by the conditional pmf $q_{UV|W}(u,v|w) = \Pr[U=u,V=v|W=w]$. The ratio of channel symbols to source symbols,
\begin{equation}
\kappa = \frac{\kappa_c}{\kappa_s}\ ,
\end{equation}
is called the bandwidth expansion. In the sequel, $\kappa_s$ and $\kappa_c$ are arbitrary fixed constants.  

A joint source-channel (JSC) blockcode of length $t$, with $\kappa_s t$ and $\kappa_c t$ being integers, is a triple of mappings $(f^{(t)}$, $g^{(t)}_1$, $g^{(t)}_2)$. Here
\begin{subequations}\label{Eqn:Enc-Dec-JSC}
\begin{equation}
f^{(t)} :\ \set{X}^{\kappa_s t} \times \set{Y}^{\kappa_s t} \rightarrow \set{W}^{\kappa_c t}\ 
\end{equation}
denotes the encoder at the transmitter, and
\begin{align}
g_1^{(t)} :&\ \set{U}^{\kappa_c t} \times \set{Y}^{\kappa_s t} \rightarrow \hat{\set{X}}^{\kappa_s t} \quad \text{ and}\\
g_2^{(t)} :&\ \set{V}^{\kappa_c t} \times \set{X}^{\kappa_s t} \rightarrow \hat{\set{Y}}^{\kappa_s t}\ .
\end{align}
\end{subequations}
denotes the decoder at receiver $i = 1,2$.

A common-reconstruction joint source-channel (CR-JSC) blockcode is a tuple of mappings $(f^{(t)}$, $g_1^{(t)}$, $g_2^{(t)}$, $\phi_1^{(t)}$, $\phi_2^{(t)})$, where $f^{(t)}$ and $g_i^{(t)}$ are defined in~\eqref{Eqn:Enc-Dec-JSC} and
\begin{subequations}\label{Eqn:CR-Enc-Dec-JSC}
\begin{align}
\phi_1^{(t)} : &\ \set{U}^{\kappa_c t} \times \set{Y}^{\kappa_s t} \rightarrow \hat{\set{Y}}^{\kappa_s t}\quad \text{ and}\\
\phi_2^{(t)} : &\ \set{V}^{\kappa_c t} \times \set{X}^{\kappa_s t} \rightarrow \hat{\set{X}}^{\kappa_s t}\ .
\end{align}
\end{subequations}
Here $\phi^{(t)}_i$ denotes the ``common-reconstruction'' decoder at receiver $i = 1,2$.

The average distortions $(\Delta_1^{(\kappa_s t)},\Delta_2^{(\kappa_s t)})$ of JSC and CR-JSC codes are defined by~\eqref{Eqn:SC-Dist-1} and~\eqref{Eqn:SC-Dist-2}, where $\kappa_s t$ replaces $n$ in the sum, and we set $\hat{\mbf{X}} \triangleq g_1^{(t)}(\mbf{U},\mbf{Y})$, $\hat{\mbf{Y}} \triangleq g_2^{(t)}(\mbf{V},\mbf{X})$ and $\mbf{W} = f^{(t)}(\mbf{X},\mbf{Y})$. The probability law of $\mbf{U}$ and $\mbf{V}$ is defined by the discrete memoryless broadcast channel
\begin{equation*}
q^{(\kappa_c t)}_{UV|W}(\mbf{u},\mbf{v}|\mbf{w}) = \prod_{i=1}^{\kappa_c t} q_{UV|W}(u_i,v_i|w_i)\ .
\end{equation*}
For the CR-JSC code, the probability of common-reconstruction decoding error $P_e$ is defined by~\eqref{Eqn:CR-Error-Prob}, where $\tilde{\mbf{Y}} \triangleq \phi_1^{(t)}(\mbf{U},\mbf{Y})$ and $\tilde{\mbf{X}} \triangleq \phi_2^{(t)}(\mbf{V},\mbf{X})$. 

\medskip

\begin{definition}[Joint Source-Channel Coding]\label{Def:JSCC}
A distortion pair $(d_1,d_2) \in \reals^2$ is said to be achievable with bandwidth expansion $\kappa$ if for every $\epsilon > 0$ there exists a joint source-channel code $(f^{(t)}$, $g_1^{(t)}$, $g_2^{(t)})$ for some sufficiently large $t$ with
\begin{equation}\label{Eqn:JSC-Distortion}
\Delta_i^{(\kappa_s t)} \leq d_i + \epsilon\ ,\quad i = 1,2.
\end{equation}
\end{definition}

\medskip

\begin{definition}[Joint Source-Channel Coding with Common-Reconstructions]\label{Def:JSCC-CR}
A distortion pair $(d_1,$ $d_2)$ $\in$ $\reals^2$ is said to be achievable with CR and bandwidth expansion $\kappa$ if for every $\epsilon > 0$ there exists a CR-JSC code $(f^{(t)}$, $g_1^{(t)}$, $g_2^{(t)}$, $\phi_1^{(t)}$, $\phi_2^{(t)})$ for some sufficiently large $t$ with $(\Delta_1^{(\kappa_s t)},\Delta_2^{(\kappa_s t)})$ satisfying~\eqref{Eqn:JSC-Distortion} and $P_e \leq \epsilon$.
\end{definition}

\subsection{Basic Rate-Distortion Functions}\label{Sec:2:BasicRDFuncts}

In this section, we briefly review some rate-distortion functions that will be used frequently throughout the paper. Let
\begin{equation}
q_X(x) \triangleq \sum_{y\in\set{Y}}q_{XY}(x,y)\ , \quad x \in \set{X}\ ,
\end{equation}
denote the $X$-marginal of $q_{XY}$. (This notation will be extended to all marginal pmfs.) Let $\set{P}_{\hat{X}|X}(d_1)$ denote the set of channels $p_{\hat{X}|X}$ mapping $\set{X}$ to $\hat{\set{X}}$ such that
\begin{equation}
\sum_{(\hat{x},x) \in \Hat{\set{X}} \times \set{X}}
p_{\hat{X}|X}(\Hat{x}|x)q_{X}(x) \delta_1(x,\Hat{x}) \leq d_1\ .
\end{equation}

\medskip

\begin{definition}[RD Function]
For $d_1 \in \reals$, the RD function of $X$ is defined by~\cite[Chap. 10]{Cover-2006-B}
\begin{equation}
R_X(d_1) \triangleq \min_{p_{\Hat{X}|X} \in \set{P}_{\hat{X}|X}(d_1)} I(X;\Hat{X})\ .
\end{equation}
\end{definition}

\medskip

Let $\set{P}_{\hat{X}\hat{Y}|XY}(d_1,d_2)$ denote the set of channels $p_{\hat{X}\hat{Y}|XY}$ mapping $\set{X} \times \set{Y}$ to $\hat{\set{X}} \times \hat{\set{Y}}$ such that
\begin{subequations}
\begin{align}
\sum_{\hat{x},\hat{y},x,y}  p_{\hat{X}\hat{Y}|XY}(\hat{x},\hat{y}|x,y) q_{XY}(x,y) \delta_1(x,\hat{x}) &\leq d_1\quad \\
\sum_{\hat{x},\hat{y},x,y}  p_{\hat{X}\hat{Y}|XY}(\hat{x},\hat{y}|x,y) q_{XY}(x,y) \delta_2(y,\hat{y}) &\leq d_2\ .
\end{align}
\end{subequations}

\medskip

\begin{definition}[Joint RD Function]\label{Def:JRD}
For $(d_1,d_2) \in \reals^2$, the joint RD function of $X$ and $Y$ is defined by~\cite{Gray-Oct-1972-A}
\begin{equation}\label{Eqn:Joint-RD}
R_{XY}(d_1,d_2) \triangleq \min _{p_{\hat{X}\hat{Y}|XY} \in \set{P}_{\hat{X}\hat{Y}|XY}(d_1,d_2)} I(X,Y;\hat{X},\hat{Y})\ .
\end{equation}
\end{definition}

\medskip

Let $\set{P}_{\hat{X}|XY}(d_1)$ denote the set of all channels $p_{\hat{X}|XY}$ mapping $\set{X} \times \set{Y}$ to $\hat{\set{X}}$ such that
\begin{equation}
\sum_{x,y,\hat{x}}p_{\hat{X}|XY}(\hat{x}|x,y) q_{XY}(x,y) \delta_1(x,\hat{x}) \leq d_1\ .
\end{equation}

\medskip

\begin{definition}[Conditional RD Function~\cite{Gray-Oct-1972-A}]\label{Def:CRD}
For $d_1 \in \reals$, the conditional RD function of $X$ given $Y$ is defined by
\begin{equation}\label{Eqn:Conditional-RD}
R_{X|Y}(d_1)\triangleq \min_{p_{\hat{X}|XY} \in \set{P}_{\hat{X}|XY}(d_1)} I(X;\hat{X}|Y)\ .
\end{equation}
\end{definition}

\medskip

Let $\set{A}$ be finite set of cardinality $|\set{A}| \leq |\set{X}|\ + 1$. Let $\set{P}_{X|Y}^{WZ}(d_1)$ denote the set of pmfs $p_{AXY}$ on $\set{A} \times \set{X} \times \set{Y}$ such that:
\begin{equation}
\sum_{a} p_{AXY}(a,x,y) = q_{XY}(x,y)\ , \ \ (x,y) \in \set{X} \times \set{Y}\ ,
\end{equation}
$A \minuso X \minuso Y$ forms a Markov chain, and there exists a function $\pi_1:\set{A}\times\set{Y}\rightarrow\Hat{\set{X}}$ such that
\begin{equation}
\sum_{(a,x,y)}p_{AXY}(a,x,y)\delta_1\big(x,\pi_1(a,y)\big)\leq d_1\ .
\end{equation}

\medskip

\begin{definition}[Wyner-Ziv RD Function]\label{Def:Wyner-Ziv}
For $d_1 \in \reals$, the Wyner-Ziv RD function for $X$ given $Y$ is defined by~\cite{Wyner-Jan-1976-A}
\begin{align}
R_{X|Y}^{WZ}(d_1) &\triangleq \min_{p \in \set{P}_{X|Y}^{WZ}(d_1)} I(X;A|Y) \ .
\end{align}
\end{definition}

\medskip

The final function that we will need to define is the minimax (or, worst noise) capacity $C_{\set{X}}(d_1)$. This function was used by Zamir in~\cite{Zamir-Nov-1996-A} to bound the rate loss in the Wyner-Ziv problem. We shall use it in a similar manner to approximate $R(d_1,d_2)$. Before defining $C_{\set{X}}(d_1)$, we first need to define the capacity of an additive channel with an input distortion constraint.

\medskip
\begin{definition}
Let $N$ be a random variable that takes values from $\set{X}=\{0,1,\ldots,l_x\}$, and let $p_{N}$ denote its pmf. Consider the additive-noise channel that randomly maps $\set{X}$ to $\set{X}$ via $x \mapsto x \oplus N$. I.e., consider $N$ to be modulo-$l_x$ additive noise. The capacity of this channel (with an input distortion constraint $d_1$) is defined by
\begin{align}
C_{\set{X}}^{add}(d_1,N) &\triangleq \sup_W I(W;W\oplus N)\ ,
\end{align}
where the supremum is taken over all choices of a random variable $W$ (defined on $\set{X}$ with pmf $p_W$ and independent of $N$) for which
\begin{equation}
\sum_{x \in \set{X}} p_{W}(x) \delta_1(x) \leq d_1\ .
\end{equation}
\end{definition}

\medskip
\begin{definition}
The minimax (worst noise) capacity under distortion constraint $d_1$ is defined by~\cite{Zamir-Nov-1996-A}
\begin{equation}
C_{\set{X}}(d_1) \triangleq \inf_{N} C_{\set{X}}^{add}(d_1,N)\ ,
\end{equation}
where the infimum is taken over all choices of a ``noise'' random variable $N$ such that
\begin{equation}
\sum_{x\in\set{X}} p_{N}(x)\delta_1(x) \leq d_1
\end{equation}
\end{definition}


\section{Main Results}\label{Sec:3}

\subsection{Main Results for Source Coding}

Our first result is a single-letter characterisation of $R_{CR}(d_1,d_2)$ for arbitrary sources and distortion measures. For $(d_1,d_2) \in \reals^2$, define
\begin{align}
R^*_{CR}(d_1,d_2) &\triangleq \min_{p_{\hat{X}\hat{Y}|XY}\in\set{P}_{\hat{X}\hat{Y}|XY}(d_1,d_2)} \max \Big\{I(X;\hat{X},\hat{Y}|Y),\ I(Y;\hat{X},\hat{Y}|X)\Big\}\ ,
\end{align}
where $\set{P}_{\hat{X}\hat{Y}|XY}(d_1,d_2)$ is defined in Section~\ref{Sec:2:BasicRDFuncts}. The next result is proved in Section~\ref{Sec:3:Proof:Thm:CR-RD}.

\medskip

\begin{theorem}\label{Thm:CR-RD}
For $(d_1,d_2) \in \reals^2$, the CR-RD function is given by
\begin{align}
R_{CR}(d_1,d_2) &= R^*_{CR}(d_1,d_2)\ .
\end{align}
\end{theorem}

\medskip

Theorem~\ref{Thm:CR-RD} is best understood in the context of the joint RD function of $X$ and $Y$. Specifically, $R^*_{CR}(d_1,d_2)$ can be rewritten as
\begin{align}
\label{Eqn:Thm:CR-RD-2}
R^*_{CR}(d_1,d_2)
&= \min_{p_{\hat{X}\hat{Y}|XY} \in \set{P}_{\hat{X}\hat{Y}|XY}(d_1,d_2)} \Big[ I(X,Y;\hat{X},\hat{Y}) - \min \big\{ I(X;\hat{X},\hat{Y}),\ I(Y;\hat{X},\hat{Y})\big\} \Big]\ ,
\end{align}
which can be interpreted as joint vector quantization coding followed by Slepian-Wolf coding. The encoder jointly maps $(\mbf{X},\mbf{Y})$ to $(\hat{\mbf{X}},\hat{\mbf{Y}})$. The common-reconstruction condition requires that $\hat{\mbf{X}}$ and $\hat{\mbf{Y}}$ satisfy the average distortion constraints $d_1$ and $d_2$, respectively. The rate needed to simultaneously satisfy these constraints is captured by the $I(X,Y;\hat{X},\hat{Y})$ term. The $\min\{I(X;\hat{X},\hat{Y}),\ I(Y;\hat{X},\hat{Y})\}$ term captures the fact that the rate $I(X,Y;\hat{X},\hat{Y})$ can be reduced by exploiting the side-information at each receiver with a Slepian-Wolf code.

\medskip

\begin{remark}
This joint vector quantization and Slepian-Wolf coding structure implicitly allows the encoder to know $\hat{\mbf{X}}$ and $\hat{\mbf{Y}}$ with high probability. We can therefore impose a third common-reconstruction constraint at the transmitter without suffering a rate-loss. That is, the RD function with common reconstructions at the transmitter and both receivers is equal to $R_{CR}(d_1,d_2)$. This result is to be expected because the transmitter has $\mbf{X}$ and $\mbf{Y}$ from which it can always compute $\hat{\mbf{X}}$ and $\hat{\mbf{Y}}$. What is less obvious, however, is that this result will also hold in the joint source-channel setting. Specifically, it will be optimal for the encoder to know $\hat{\mbf{X}}$ and $\hat{\mbf{Y}}$ with high probability. This result is not obvious because it is sometimes necessary to exploit randomness in the channel to efficiently induce distortions~\cite{Gastpar-Nov-2008-A}.
\end{remark}

\medskip

Theorem~\ref{Thm:CR-RD} gives a relatively straightforward single-letter characterisation of $R_{CR}(d_1,d_2)$. 
In contrast, giving a single-letter characterisation of $R(d_1,d_2)$ is much more difficult. A simple lower bound for $R(d_1,d_2)$  stems from the following cut-set argument: $R(d_1,d_2)$ must be at least as large as the smallest rate that is needed to compress $\mbf{X}$ at the transmitter for decoding by receiver $1$, while ignoring the distortion constraint on $\mbf{Y}$ for receiver $2$. The smallest such rate is given by the conditional RD function $R_{X|Y}(d_1)$. More formally, we have the following. 

\medskip

\begin{proposition}\label{Prop:Cut-Set}
For $(d_1,d_2) \in \reals^2$, we have that
\begin{equation}
R(d_1,d_2) \geq R_L(d_1,d_2) \ ,
\end{equation}
where 
\begin{equation}
R_L(d_1,d_2) \triangleq  \max\big\{R_{X|Y}(d_1), R_{Y|X}(d_2)\big\}\ .
\end{equation}
\end{proposition}

\medskip

Surprisingly, $R_L(d_1,d_2)$ is the tightest lower bound in the literature. It equals $R(d_1,d_2)$ in the high-distortion regime where $d_1 = d_{1,\text{max}}$ or $d_2 = d_{2,\text{max}}$, but it is an open problem as to whether $R_L(d_1,d_2)$ always equals\footnote{Two upper bounds for $R(d_1,d_2)$ have been given in~\cite{Su-Jun-2010-C} and~\cite{Kimura-Apr-2008-A}. We discuss these bounds in Section~\ref{Sec:4}.} $R(d_1,d_2)$. The next example describes a simple binary source where $R_L(d_1,d_2)$ is equal to $R(d_1,d_2)$. This example was also given in~\cite{Su-Jun-2010-C}. We review it here because it is relevant to the following discussion.

\medskip

\begin{definition}\label{Def:DSBS}
The source $q_{XY}$ is said to be a Doubly Symmetric Binary Source (DSBS) with cross-over probability $\rho$ if $\set{X}$ $=$ $\hat{\set{X}}$ $=$ $\set{Y}$ $=$ $\hat{\set{Y}}$ $=$ $\{0,1\}$, $\rho \in [0,1/2]$ and
\begin{equation}
q_{XY}(x,y) \triangleq \frac{1}{2}(1-\rho) \mathbf{1}_{x,y} + \frac{1}{2}\rho(1-\mathbf{1}_{x,y})\ ,
\end{equation}
where
\begin{equation}
\mathbf{1}_{x,y} \triangleq \left\{
                              \begin{array}{ll}
                                0, & \hbox{ if } x = y\\
                                1, & \hbox{ otherwise. }
                              \end{array}
                            \right.
\end{equation}
We can view $q_{XY}$ as resulting from the equation $Y = X \oplus Z$. Here $X$ is uniform on $\set{X}$, $\oplus$ denotes modulo-two addition, and $Z$ is independent of $X$ and takes values from $\{0,1\}$ with probability $q_Z(0) = 1 - \rho$ and $q_Z(1) = \rho$.
\end{definition}

\medskip

\begin{example}\label{Exa:DSBS}
If $q_{XY}$ is the DSBS with cross-over probability $\rho$ and $\delta_1$ and $\delta_2$ are Hamming measures, then for all $d \in [0,1]$ we have that~\cite{Gray-Oct-1972-A}
\begin{equation}
R_{X|Y}(d)
= R_{Y|X}(d)
= \left\{
\begin{array}{ll}
h(\rho) - h(d), & \hbox{ if } d \leq \rho\ ,\\
0, & \hbox{ otherwise,}
\end{array}
\right.
\end{equation}
where
\begin{equation}
h(\lambda) \triangleq -\lambda\log_2\lambda -(1-\lambda)\log_2(1-\lambda)
\end{equation}
is the binary entropy function (take $h(0) = h(1) = 0$). Let $d_{\text{min}} \triangleq \min \{d_1,d_2\}$. Clearly, we have that $R(d_1,d_2) = R_L(d_1,d_2) = 0$ for $d_{\text{min}} \geq \rho$ because each receiver can estimate its reconstruction directly from its side-information. For $d_{\text{min}} < \rho$, the transmitter computes $\mbf{Z} = \mbf{X} \oplus \mbf{Y}$ and sends a distorted version $\hat{\mbf{Z}}$ of $\mbf{Z}$ to both receivers with an average (Hamming) distortion of $d_{\text{min}}$. This can be done with a binary RD code of rate $R_Z(d_\text{min}) = h(\rho) - h(d_{\text{min}})$; for example, see~\cite[Thm. 10.3.1]{Cover-2006-B}. Receiver $1$ decodes $\hat{\mbf{X}}$ by setting $\hat{X}_i = \hat{Z}_i \oplus Y_i$ for $i = 1,2,\ldots,n$. Similarly, receiver $2$ decodes $\hat{\mbf{Y}}$ by setting $\hat{Y}_i = \hat{Z}_i \oplus X_i$. It can be verified that both reconstructions, $\hat{\mbf{X}}$ and $\hat{\mbf{Y}}$, achieve an average distortion $d_{\text{min}}$. The RD function is therefore given by
\begin{equation}
R(d_1,d_2) =
\left\{
\begin{array}{ll}
h(\rho) - h(d_{\text{min}}), & \hbox{ if } d_{\text{min}} \leq \rho \\
0, & \hbox{ otherwise.}
\end{array}
\right.
\end{equation}
It is worth noting that the above code achieves an average distortion $d_{\text{min}}$ for both receivers; that is, it operates at the point $R(d_{\text{min}},d_{\text{min}})$. Note also that this code does not satisfy Definition~\ref{Def:CR-Source-Coding} (e.g., receiver $1$ cannot compute $\hat{Y}_i = \hat{Z}_i \oplus X_i$), so it cannot be used as a CR-RD code. The RD function is plotted for $\rho = 0.25$ in Figure~\ref{Fig:RLDSBS}.
\begin{figure}[t]
\begin{center}
\includegraphics[width=65mm]{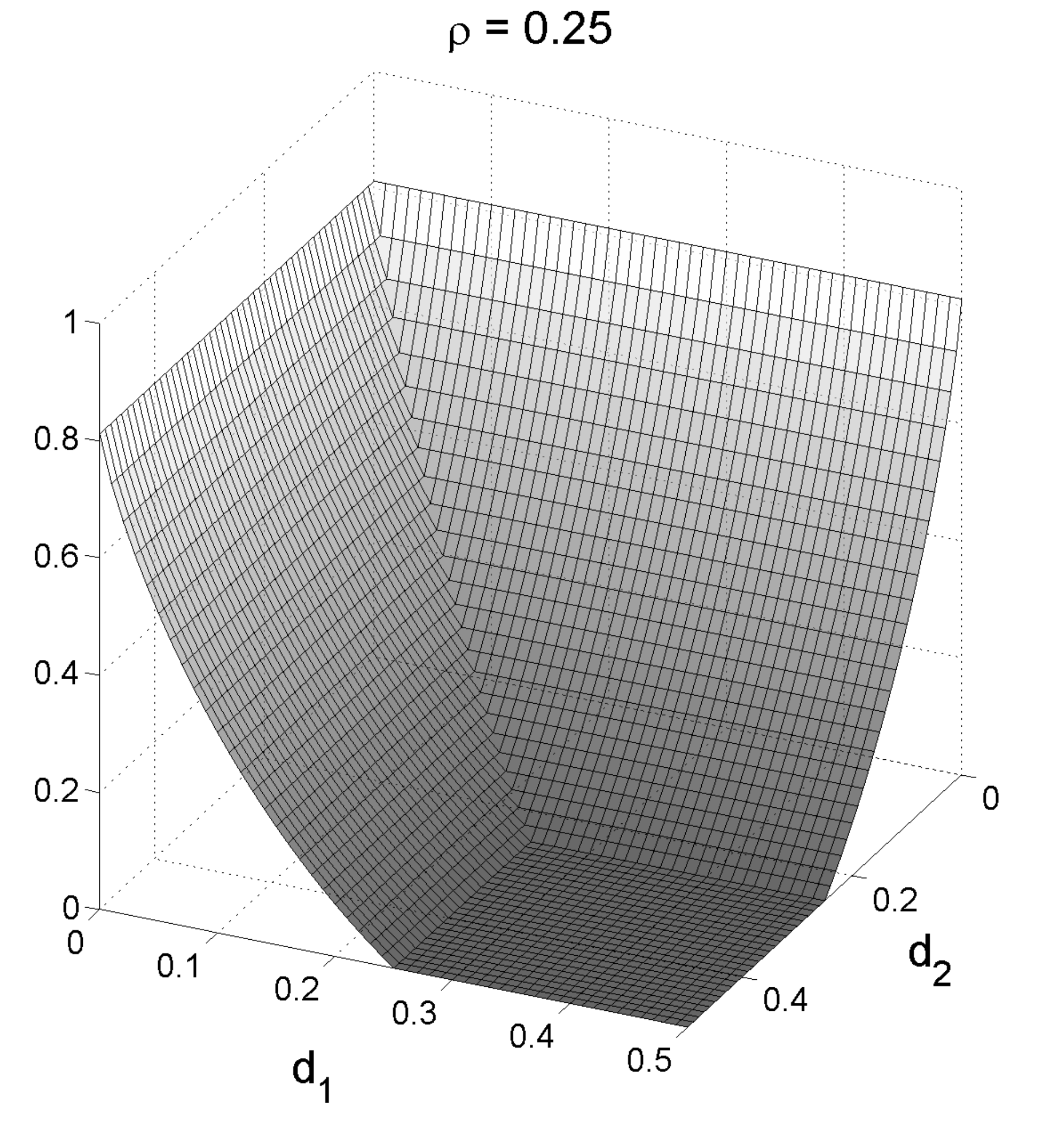}\\
\end{center}
  \caption{Figure shows the RD function $R(d_1,d_2) = h(\rho) - h(d_{\text{min}})$ for the doubly symmetric binary source (DSBS) with cross over probability $\rho = 0.25$ and Hamming distortions. This function is developed in Example~\ref{Exa:DSBS}.}\label{Fig:RLDSBS}
\end{figure}
\end{example}

\medskip

Consider the three functions: the CR-RD function $R_{CR}(d_1,d_2)$, the RD function $R(d_1,d_2)$ and the cut-set lower bound $R_L(d_1,d_2)$. It is clear that
\begin{equation}\label{Eqn:RD-CR-RD}
R_{CR}(d_1,d_2) \mathop{\geq}^{(a)} R(d_1,d_2) \mathop{\geq}^{(b)} R_L(d_1,d_2)\ ,\quad  (d_1,d_2) \in \reals^2\ ,
\end{equation}
for all sources and distortion measures. Inequality $(a)$ can be strict. For example, in Example~\ref{Exa:DSBS} there is zero common information (in the G\'{a}cs-K\"{o}rner~\cite{Gacs-1973-A} sense) between $X$ and $Y$ when $\rho > 0$. This means that the receivers cannot agree on any non-trivial $\hat{\mbf{X}}$ and $\hat{\mbf{Y}}$ without additional information from the transmitter. Therefore, one would expect that $R_{CR}(d,d)$ cannot reach $0$ until $d = 0.5$. In contrast, note that $R(d,d) = 0$ for all $d \geq \rho$ because each receiver can estimate its reconstruction directly from its side-information; see, for example, $d_1 = d_2 = 0.25$ in Figure~\ref{Fig:RLDSBS}.

The next result shows that both $(a)$ and $(b)$ are equalities for vanishing Hamming distortions. The proof involves a minor modification of a result by Sgarro~\cite{Sgarro-Mar-1977-A} (see also Wyner {\em et al.}~\cite[Thm. 1]{Wyner-Jun-2002-A}) and is omitted for brevity.

\medskip

\begin{proposition}[Sgarro~\cite{Sgarro-Mar-1977-A}]\label{Prop:LosslessSC}
If $\delta_1$ and $\delta_2$ are Hamming distortion measures, then
\begin{equation}\label{Eqn:Sgarro}
R_{CR}(0,0) = R(0,0) = \max\big\{H(X|Y),H(Y|X)\}\ .
\end{equation}
\end{proposition}

\medskip

Our next result shows that inequalities $(a)$ and $(b)$ are in fact equalities for a non-trivial range of small distortions. A surface $\set{D}$ in $\reals^2$ is said to be strictly positive if for all $(d_1,d_2) \in \set{D}$ we have $d_1 > 0$ and $d_2 > 0$; see, for example, Gray~\cite{Gray-Jul-1973-A}. The next result is proved in Section~\ref{Sec:3:Small-Distortions}.

\medskip

\begin{theorem}\label{Thm:Small-Distortions}
If $q_{XY}$ has support $\set{X} \times \set{Y}$ and $\delta_1$ and $\delta_2$ are Hamming distortion measures, then there exists a strictly positive surface $\set{D}$ in $\reals^2$ such that 
\begin{equation}
R_{CR}(d_1,d_2) = R(d_1,d_2) = R_L(d_1,d_2) \equiv \max \big\{R_{X|Y}(d_1),R_{Y|X}(d_2) \big\}\ ,
\end{equation}
whenever $(d_1,d_2)$ lies on or below $\set{D}$; that is, there exists some $(d_1',d_2') \in \set{D}$ with $d_1\leq d_1'$ and $d_2 \leq d_2'$.
\end{theorem}

\medskip

This result is not just interesting because $R(d_1,d_2)$ and $R_{CR}(d_1,d_2)$ both meet the cut-set lower bound $R_L(d_1,d_2)$ for small distortions. It also gives an explicit characterisation of $R(d_1,d_2)$ for a class of sources and distortions for which $R(d_1,d_2)$ would be otherwise unknown. 

We prove Theorem~\ref{Thm:Small-Distortions} by matching the cut-set lower bound $R_L(d_1,d_2)$ to the single-letter characterisation of the CR-RD function $R_{CR}(d_1,d_2)$ given in Theorem~\ref{Thm:CR-RD}. An important step in this proof requires that Gray's extended Shannon lower bounds for joint, conditional and marginal RD functions~\cite{Gray-Jul-1973-A} are tight. This tightness is only achieved in the small distortion regime\footnote{We note in passing that Shannon lower bounds are often used to prove small-distortion results; for example, see~\cite{Linder-Nov-1994-A,Zamir-Jan-1999-A,Zamir-Nov-1999-A}.}.

The notion of ``small distortions'' is not vacuous; our next result shows that the set of distortions for which Theorem~\ref{Thm:Small-Distortions} holds for the DSBS is in fact quite large. Moreover, the boundary of this set has a close connection to common information (in Wyner's sense~\cite{Wyner-Mar-1975-A}). Let $\set{W}$ be a finite set of cardinality $|\set{W}| \leq 4$ and let~\cite{Wyner-Mar-1975-A}
\begin{equation}
K(X;Y) \triangleq \min_{p_{W|XY} \in \set{P}_{W|XY}} I(X,Y;W)\ ,
\end{equation}
where $\set{P}_{W|XY}$ is the set of channels $p_{W|XY}$ mapping $\set{X}\times\set{Y}$ to $\set{W}$ such that the resulting joint pmf for $(X,Y,W)$ forms the Markov chain $X \minuso W \minuso Y$. The next result is proved in Section~\ref{Sec:3:DSBS}.

\medskip

\begin{theorem}\label{Thm:DSBS-CR}
If $q_{XY}$ is the DSBS with cross-over probability $\rho \in [0,1/2]$, $\delta_1$ and $\delta_2$ are Hamming distortion measures, and
\begin{equation}
d^* \triangleq \frac{1}{2} - \frac{1}{2} \sqrt{1-2\rho}\ ,
\end{equation}
then the CR-RD function $R_{CR}(d,d)$ satisfies the following:
\begin{enumerate}
\item[(i)] For all $d \in [0,d^*]$
\begin{align}\label{Thm:DSBS-CR:Part1}
R_{CR}(d,d) = R(d,d) = h(\rho) - h(d)\ ;
\end{align}
\item[(ii)]
\begin{subequations}\label{Thm:DSBS-CR:Part2}
\begin{align}
R_{CR}(d^*,d^*) &= K(X;Y) - R_X(d^*)\\
& = K(X;Y) - R_Y(d^*)\ ;
\end{align}
\end{subequations}
\item[(iii)] For all $d \in (d^*,1/2]$
\begin{subequations}\label{Thm:DSBS-CR:Part3}
\begin{align}
\label{Thm:DSBS-CR:Part3a}
R_{CR}(d,d) &\neq h(\rho) - h(d)\ , \text{ and}\\
\label{Thm:DSBS-CR:Part3b}
R_{CR}(d,d) &\leq h(d) - \rho - (1-\rho)h\left(\frac{2d-\rho}{2(1-\rho)}\right)\ .
\end{align}
\end{subequations}
\end{enumerate}
\end{theorem}

\begin{figure}[h!]
\centering
\subfigure[]{
\includegraphics[width=50mm]{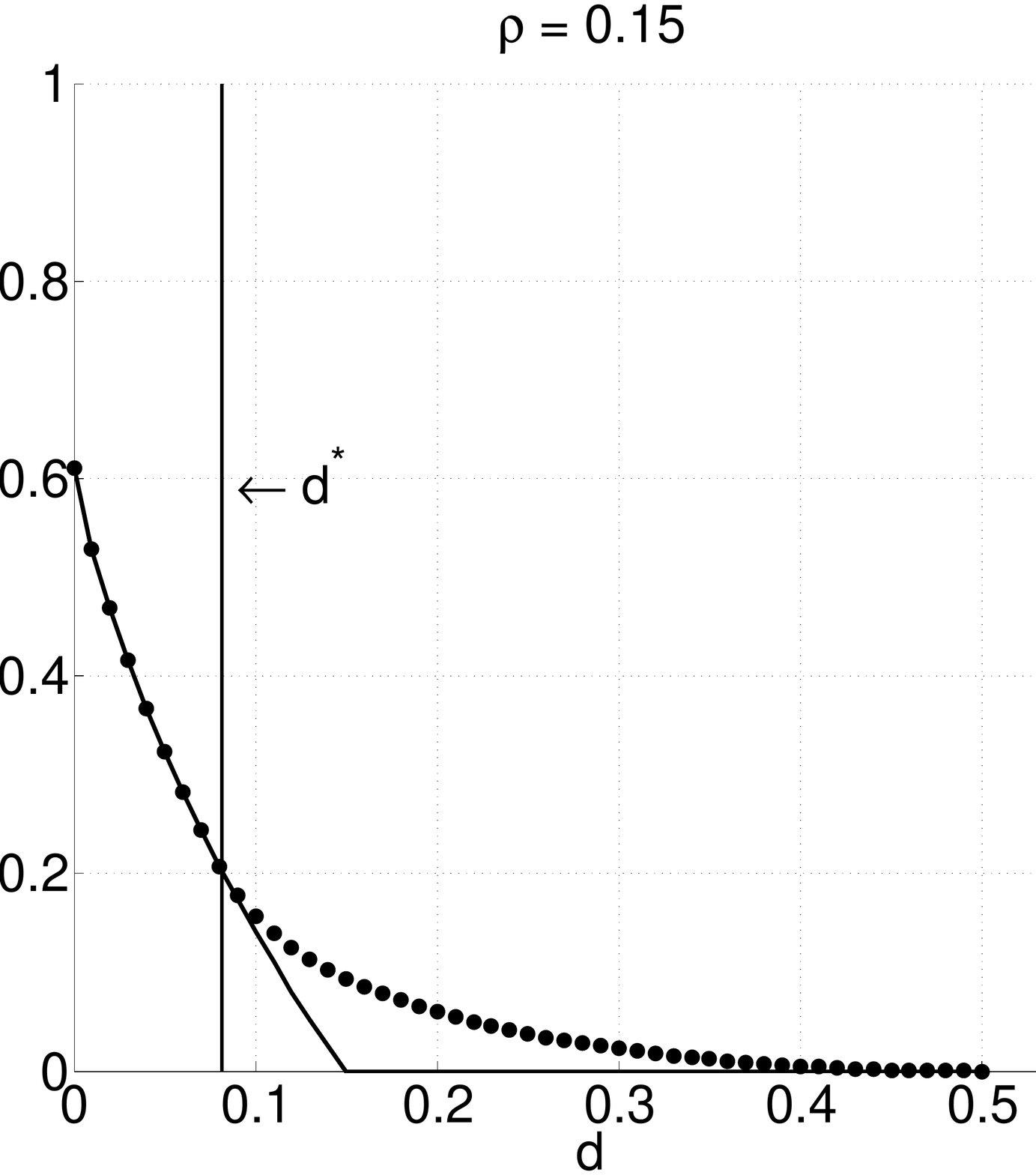}
\label{Fig:DSBS-1a}
}
\subfigure[]{
\includegraphics[width=50mm]{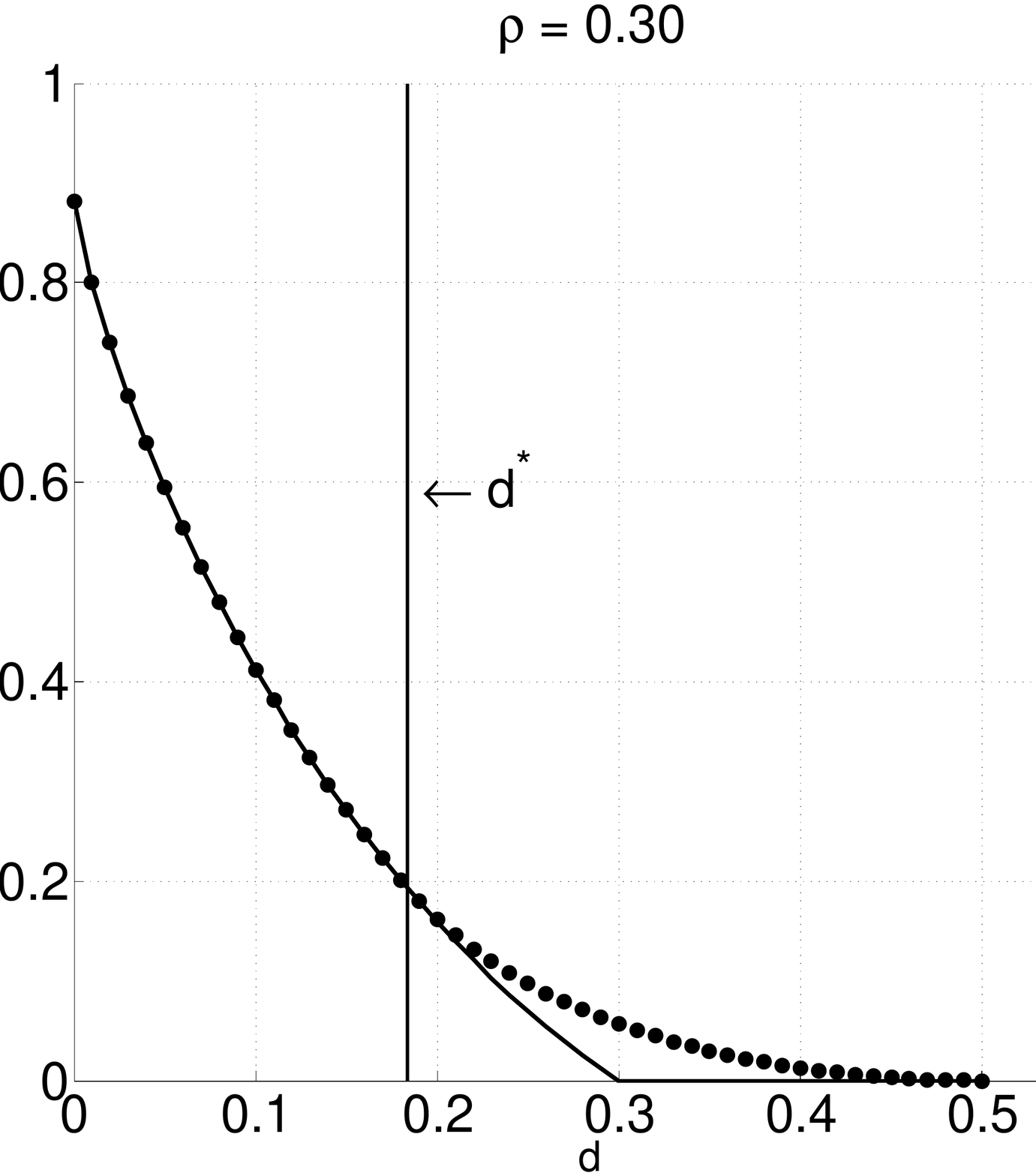}
\label{Fig:DSBS-1b}
}
\subfigure[]{
\includegraphics[width=50mm]{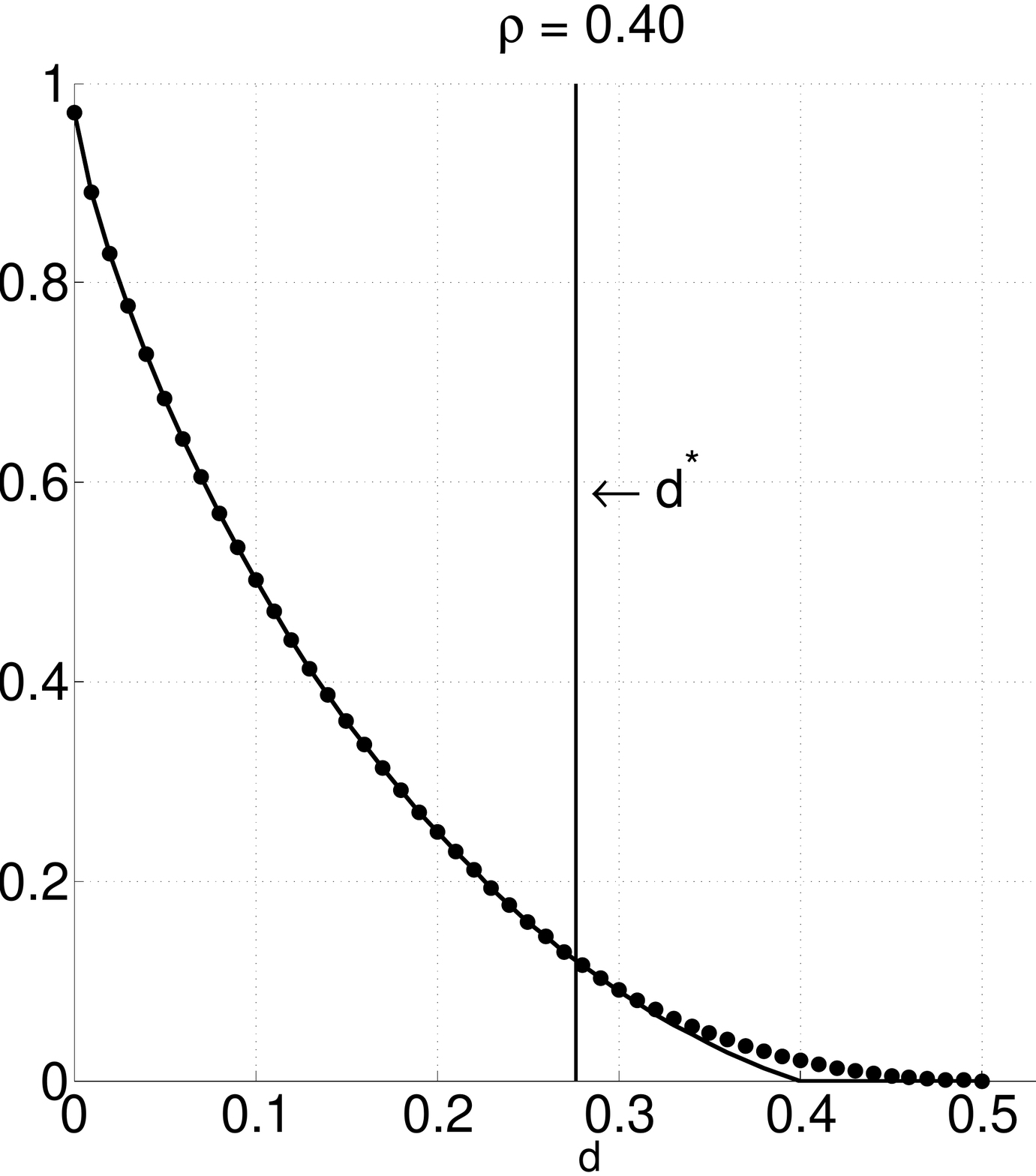}
\label{Fig:DSBS-1c}
}
\caption{The RD function $R(d_1,d_2)$ as well as an upper bound for the CR-RD function $R_{CR}(d_1,d_2)$ are plotted for the DSBS with cross-over probability $\rho$. We consider three different values $\rho$. In Figure~\ref{Fig:DSBS-1a} we have $\rho = 0.15$, in Figure~\ref{Fig:DSBS-1b} we have $\rho = 0.30$ and in Figure~\ref{Fig:DSBS-1c} we have $\rho = 0.40$. The RD function $R(d_1,d_2)$ is identified by a solid line, the upper bound for $R_{CR}(d_1,d_2)$ is identified by by a dotted line, and the threshold $d^*$ is identified by a vertical solid line. In all three plots we have set $d_1 = d_2 = d$.}
\label{Fig:DSBS-1}
\end{figure}

In Figure~\ref{Fig:DSBS-1} we plot $R(d,d)$, $d^*$, and the upper bound for $R_{CR}(d,d)$ that is given in~\eqref{Thm:DSBS-CR:Part3b}. It can be seen from these plots that the threshold $d^*$ is reasonably large, and most interesting distortion pairs can be achieved by a CR-RD code.

\subsection{Main Results for Joint Source-Channel Coding}

Our next result characterises joint source-channel coding rates with common reconstructions. It is the joint source-channel coding extension of the Theorem~\ref{Thm:CR-RD}.

\medskip

\begin{theorem}\label{Thm:JSCC-CR}
A distortion pair $(d_1,d_2) \in \reals^2$ is achievable with common reconstructions and bandwidth expansion $\kappa$ if and only if there exists a pmf $p_W$ on $\set{W}$ and $p_{\hat{X}\hat{Y}|XY}$ $\in$ $\set{P}_{\hat{X}\hat{Y}|XY}(d_1,d_2)$ such that
\begin{subequations}\label{Eqn:Thm:JSCC-CR-1}
\begin{align}
I(X;\hat{X},\hat{Y}|Y) &\leq \kappa I(W;U)\ \\
I(Y;\hat{X},\hat{Y}|X) &\leq \kappa I(W;V)\ .
\end{align}
\end{subequations}
\end{theorem}
\medskip

As was the case for source coding, characterising joint source-channel coding rates without common-reconstructions (i.e. Definition~\ref{Def:JSCC}) is difficult, and we have succeeded only in giving complete results for a few special cases. The next proposition reviews a special case that is known in the literature. This proposition follows from Tuncel~\cite[Thm. 6]{Tuncel-Apr-2006-A}, and it can be thought of as the joint source-channel coding extension of Sgarro's result (Proposition~\ref{Prop:LosslessSC}).

\medskip

\begin{proposition}[Tuncel~\cite{Tuncel-Apr-2006-A}]\label{Prop:LosslessJSCC}
Suppose $\delta_1$ and $\delta_2$ are Hamming distortion measures. Zero distortion is achievable with bandwidth expansion $\kappa$ if and only if there exists a pmf $p_W$ on $\set{W}$ such that
\begin{subequations}
\label{Eqn:Tuncel}
\begin{align}
H(X|Y) &\leq \kappa\ I(W;U)\quad \text{and}\\
H(Y|X) &\leq \kappa\ I(W;V)\ .
\end{align}
\end{subequations}
\end{proposition}
\medskip

Tuncel's result is ideal because it characterises achievability simply and explicitly; it does not require auxiliary random variables and difficult optimization problems to be solved. The following consequences of this result are worth noting: $(i)$ the physical separation of source and channel codes is suboptimal\footnote{When considering separate source and channel codes, Tuncel~\cite{Tuncel-Apr-2006-A} assumed that the side-information present at each receiver is not used in the channel code. This assumption is appropriate in~\cite{Tuncel-Apr-2006-A} because the side-information can be arbitrarily distributed. However, in Proposition~\ref{Prop:LosslessJSCC} the side-information takes a particular ``complimentary'' form, and in some circumstances it may be appropriate to use this side-information in the channel code; for example, see~\cite{Timo-Nov-2010-T}.}; $(ii)$ an optimal joint source-channel code exhibits a ``partial'' separation of source and channel coding at the transmitter, which results in the separation of source and channel random variables in~\eqref{Eqn:Tuncel}; $(iii)$ an optimal joint source-channel code exploits randomness in the broadcast channel to perform a ``virtual binning,'' which is analogous to the random binning used in the proof of Proposition~\ref{Prop:LosslessSC}; $(iv)$ if the broadcast channel is such that the same $p_W$ maximises $I(W;U)$ and $I(W;V)$, then all channels can be used to full capacity. This last property is not shared by broadcast channels in general.

Like Sgarro's result for lossless source coding (Proposition~\ref{Prop:LosslessSC}), Tuncel's result does not easily extend to more general distortion measures and distortions. This difficulty is evidenced by the growing body of work~\cite{Gunduz-Jul-2008-C,Nayak-Apr-2010-A1,Wilson-Feb-2008-arxiv,Behroozi-Oct-2009-C,Gao-2010-A} concerning the lossy extension of~\cite{Tuncel-Apr-2006-A}. Our next result gives necessary conditions for a distortion pair to be achievable. It is the joint source-channel coding extension of the cut-set lower bound $R_L(d_1,d_2)$ for $R(d_1,d_2)$, see Proposition~\ref{Prop:Cut-Set}. A proof of this result is given in Section~\ref{Sec:4}.

\medskip

\begin{theorem}\label{Thm:JSCC-Cut-Set}
If $(d_1,d_2) \in \reals^2$ is achievable with bandwidth expansion $\kappa$, then there exists a pmf $p_W$ on $\set{W}$ such that
\begin{subequations}\label{Eqn:ThmJSCC-Cut-Set}
\label{Eqn:JSCC-Cut-Set}
\begin{align}
\label{Eqn:JSCC-Cut-Set-1}
R_{X|Y}(d_1) &\leq \kappa\ I(W;U)\ \text{ and}\\
\label{Eqn:JSCC-Cut-Set-2}
R_{Y|X}(d_2) &\leq \kappa\ I(W;V)\ .
\end{align}
\end{subequations}
\end{theorem}

\medskip

In the Hamming distortion setting, we have that $R_{X|Y}(0) = H(X|Y)$ and $R_{Y|X}(0) = H(Y|X)$. Therefore,  Theorem~\ref{Thm:JSCC-Cut-Set} gives the necessary (``only if'') condition of Proposition~\ref{Prop:LosslessJSCC}. Similarly, in the high-distortion regime $d_2 = d_{2,\text{max}}$ we have that $R_{Y|X}(d_2) = 0$ and~\eqref{Eqn:JSCC-Cut-Set-2} is satisfied by any $p_W$. We are left with~\eqref{Eqn:JSCC-Cut-Set-1}, which is the necessary condition of Shannon's joint source-channel coding theorem~\cite[Thm. 9.2.2]{Gallager-1968-B}. It is an open problem as to whether the conditions of Theorem~\ref{Thm:JSCC-Cut-Set} are both necessary and sufficient. The next result shows that these conditions are necessary and sufficient for small distortions.

\medskip

\begin{theorem}\label{Thm:JSCC-Small-Distortions}
Suppose $q_{XY}$ has support $\set{X} \times \set{Y}$ and $\delta_1$ and $\delta_2$ are Hamming distortion measures. There exists a strictly positive surface $\set{D}$ in $\reals^2$ such that every $(d_1,d_2)$ on or below $\set{D}$ is achievable with bandwidth expansion $\kappa$ if and only if there exists a pmf $p_W$ on $\set{W}$ such that~\eqref{Eqn:JSCC-Cut-Set} holds.
\end{theorem}

\medskip

The proof of Theorem~\ref{Thm:JSCC-Small-Distortions} follows in a similar manner to the proof of Theorem~\ref{Thm:Small-Distortions}. Specifically, we match the single-letter characterisation of Theorem~\ref{Thm:JSCC-CR} with the necessary conditions in Theorem~\ref{Thm:JSCC-Cut-Set}.


\section{Source Coding: Auxiliary Results $\&$ Proofs}\label{Sec:4}

\subsection{Approximating $R(d_1,d_2)$}\label{Sec:4A}

We have already reviewed the cut-set lower bound
\begin{equation}
R(d_1,d_2) \geq R_L(d_1,d_2) \equiv \max \big\{R_{X|Y}(d_1),R_{Y|X}(d_2)\big\}
\end{equation}
in the introduction. We now review an upper bound for $R(d_1,d_2)$ that, together with $R_L(d_1,d_2)$, gives a good approximation of $R(d_1,d_2)$.

Let
\begin{equation}\label{Eqn:RD-CL-Upper-Bound}
R_{U}(d_1,d_2) \triangleq \max \big\{ R_{X|Y}^{WZ}(d_1),R_{Y|X}^{WZ}(d_2)\big\}\ .
\end{equation}
Su and El. Gamal~\cite{Su-Jun-2010-C} called this bound the compress-linear upper bound~\cite{Su-Jun-2010-C} -- the reason will become clear shortly. If $\delta_1$ and $\delta_2$ are difference distortion measures, let
\begin{equation}
C(d_1,d_2) \triangleq \max\big\{ C_{\set{X}}(d_1),\ C_{\set{Y}}(d_2) \big\}\ .
\end{equation}

The next result bounds $R(d_1,d_2)$ from above and below, and it approximates $R(d_1,d_2)$ when $d_1$ and $d_2$ are difference distortion measures.

\medskip

\begin{theorem}\label{Thm:RD-Bounds}
For $(d_1,d_2) \in \reals^2$, we have that~\cite[Thm. 2]{Su-Jun-2010-C}
\begin{equation}\label{Eqn:RD-Upper-Lower-Bound}
R_{L}(d_1,d_2) \leq R(d_1,d_2) \leq R_{U}(d_1,d_2)\ .
\end{equation}
If $\delta_1$ and $\delta_2$ are difference distortion measures, then 
\begin{equation}\label{Eqn:RD-Approximation}
R_{U}(d_1,d_2) - R_{L}(d_1,d_2) \leq C(d_1,d_2)\ .
\end{equation}
\end{theorem}

\medskip

The minimax capacity bound~\eqref{Eqn:RD-Approximation} shows that the gap between $R_L(d_1,d_2)$ and $R_U(d_1,d_2)$ cannot be arbitrarily large~\cite{Zamir-Nov-1996-A}. The inequalities in~\eqref{Eqn:RD-Upper-Lower-Bound} were obtained independently and contemporaneously by Su and El. Gamal in~\cite{Su-Jun-2010-C}. This proof of Theorem~\ref{Thm:RD-Bounds} is relevant to the following discussion, so it is worthwhile to give a brief outline.

\medskip

\begin{proof}
The fact that $R(d_1,d_2) \geq R_L(d_1,d_2)$ follows from the cut-set argument given in the introduction. To show $R(d_1,d_2) \leq R_U(d_1,d_2)$ we combine two Wyner-Ziv codes with a simple linear-network code. At the transmitter, $\mbf{X}$ is mapped to a binary vector using an optimal Wyner-Ziv code~\cite{Wyner-Jan-1976-A}. This code treats $\mbf{Y}$ as side-information at receiver $1$, but it ignores $\mbf{Y}$ at the transmitter. Similarly, $\mbf{Y}$ is mapped to a binary vector using a Wyner-Ziv code that treats $\mbf{X}$ as side-information at receiver $2$, but it ignores $\mbf{X}$ at the transmitter. The transmitter sends the modulo-two sum of these codewords (in the same way as Example~\ref{Exa:DSBS}) over the noiseless BC, and each receiver recovers their desired codeword by eliminating (subtracting) the codeword destined for the other receiver. It is possible to perform this elimination
because each receiver can calculate (from its side-information) the Wyner-Ziv codeword intended for the other receiver. Note, if conditional RD codes were used in place of Wyner-Ziv codes, then each receiver cannot calculate the codeword intended for the other user and this elimination is not possible. The second result~\eqref{Eqn:RD-Approximation} follows directly from Zamir's work on rate-loss in the Wyner-Ziv problem~\cite{Zamir-Nov-1996-A}. $R_U(d_1,d_2)$ is called the compress-linear upper bound because it is obtained by combining two Wyner-Ziv compression codes with a linear-network code.
\end{proof}

\medskip

The gap between $R_{U}(d_1,d_2)$ and $R_{L}(d_1,d_2)$ can be no larger than the ``rate loss'' of the Wyner-Ziv RD function over the conditional RD function. If $q_{XY}$ and $\delta_1$ and $\delta_2$ are such that there is no rate loss, then Theorem~\ref{Thm:RD-Bounds} characterises $R(d_1,d_2)$. The following examples outline a number of such scenarios.

\medskip

\begin{corollary}[Conditional Independence]\label{Thm:RD-Bounds:Cor:Cond-Ind}
If $X = (X',U)$ and $Y=(Y',U)$ where $X' \minuso U \minuso Y'$ forms a Markov chain, then for all distortion pairs $(d_1,d_2) \in \reals^2$ we have that
\begin{align}
R(d_1,d_2) &= R_L(d_1,d_2)\\
&= R_U(d_1,d_2)\\
&= \max\big\{R_{X|U}(d_1),R_{Y|U}(d_2)\big\}\ .
\end{align}
In particular, if $X$ and $Y$ are independent, then we have 
\begin{equation}
R(d_1,d_2) = \max\big\{R_{X}(d_1),R_{Y}(d_2)\big\}\ .
\end{equation}
\end{corollary}

\medskip

\begin{proof}
If $X = (X',U)$ and $Y=(Y',U)$ where $X' \minuso U \minuso Y'$ forms a Markov chain, then $X \minuso U \minuso Y$ also forms a Markov chain. Moreover, we have
\begin{equation}\label{Eqn:The:RD-Bounds:Cor:Cond-Ind-1}
R_{X|Y}(d_1) \mathop{\geq}^{(a)} R_{X|U}(d_1) \mathop{=}^{(b)} R^{WZ}_{X|U}(d_1) \mathop{\geq}^{(c)} R^{WZ}_{X|Y}(d_1)\ ,
\end{equation}
where $(a)$ follows from the Markov chain $X \minuso U \minuso Y$, $(b)$ follows because\footnote{The side-information $U$ is a component of the source; therefore, $R_{X|U}^{WZ}(d_1)$ and $R_{X|U}(d_1)$ are equal.} $X = (X',U)$, and $(c)$ follows because $Y = (Y',U)$. On combining~\eqref{Eqn:The:RD-Bounds:Cor:Cond-Ind-1} with the fact that $R^{WZ}_{X|Y}(d_1) \geq R_{X|Y}(d_1)$, it follows that $R^{WZ}_{X|Y}(d_1) = R_{X|Y}(d_1)$. A similar argument yields $R^{WZ}_{Y|X}(d_2) = R_{Y|X}(d_2)$. Substituting these equalities into the definitions of $R_L(d_1,d_2)$ and $R_U(d_1,d_2)$, and applying Theorem~\ref{Thm:RD-Bounds} completes the proof.
\end{proof}

\medskip

\begin{corollary}[Two Deterministic Reconstructions]\label{Thm:RD-Bounds:Cor:2Deterministic}
If $\set{U}$ and $\set{V}$ are finite sets, $\psi_x : \set{X} \rightarrow \set{U}$ and $\psi_y : \set{Y} \rightarrow \set{V}$ are mappings, $U = \psi_x(X)$, $V = \psi_y(Y)$,
$\Hat{\set{X}} = \set{U}$, $\Hat{\set{Y}} = \set{V}$,
\begin{align}
\label{Eqn:Deterministic-Distortion-X}
\delta_1(x,\Hat{u}) &\triangleq\left\{
                 \begin{array}{ll}
                   0, & \hbox{ if } u = \psi_x(x) \\
                   1, & \hbox{ otherwise,}
                 \end{array}
               \right.\\
\label{Eqn:Deterministic-Distortion-Y}
\delta_2(y,\Hat{v}) &\triangleq\left\{
                 \begin{array}{ll}
                   0, & \hbox{ if } v = \psi_y(y) \\
                   1, & \hbox{ otherwise,}
                 \end{array}
               \right.
\end{align}
then we have that
\begin{equation}
R(0,0) = \max\big\{H(U|Y),H(V|X)\big\}\ .
\end{equation}
\end{corollary}

\medskip

\begin{proof}
The conditional RD function $R_{X|Y}(d_1)$ and the Wyner-Ziv RD function $R_{X|Y}^{WZ}(d_1)$ are both continuous\footnote{The Wyner-Ziv rate-distortion function was shown to be continuous at $d = 0$ by Willems in~\cite{Willems-Jul-1983-T}. The continuity of the conditional rate distortion function $R_{X|Y}(d_1)$ at $d_1 = 0$ follows from Willems result because $R_{X|Y}(d_1)$ is a special case of the Wyner-Ziv rate distortion function when the source and distortion measure are chosen appropriately.} at $d_1 = 0$. We have that
\begin{equation}
R_{X|Y}(0) = \min I(X;\Hat{U}|Y)\ ,
\end{equation}
where the minimum is taken over all channels $p_{\hat{U}|XY}$ with
\begin{equation}\label{Eqn:RD-Bounds:Cor:2Deterministic-2}
\sum_{(\hat{u},x,y)\in\set{U}\times\set{X}\times\set{Y}} p_{\hat{U}|XY}(\hat{u}|x,y)q_{XY}(x,y) \delta_1(x,\hat{u}) = 0\ .
\end{equation}
Suppose that $p_{\hat{U}|XY}$ achieves the above minimum. Since $\delta_1(x,\hat{u}) = 0$ when $\psi_x(x) = \hat{u}$ and $\delta_1(x,\hat{u}) = 1$ when $\psi_x(x) \neq \hat{u}$, \eqref{Eqn:RD-Bounds:Cor:2Deterministic-2} implies that when $q_{XY}(x,y) > 0$ we have that $p_{\hat{U}|XY}$ must satisfy
\begin{align}\label{Eqn:RD-Bounds:Cor:2Deterministic-4}
p_{\hat{U}|XY}(\hat{u}|x,y) &=
\left\{
\begin{array}{ll}
  1, & \hbox{ if } \hat{u} = \psi_x(x)\\
  0, & \hbox{ otherwise.}
\end{array}
\right.
\end{align}
That is, $\hat{U} = U$ almost surely. Therefore $H(\hat{U}|X,Y) = 0$ and $R_{X|Y}(0) = H(\hat{U}|Y) = H(U|Y)$. We also have that
\begin{equation}
R_{X|Y}^{WZ}(d_1) = \min I(X;A|Y)\ ,
\end{equation}
where the minimization is taken over all choices of an auxiliary random variable $A$ with a joint pmf $p_{AXY}$ satisfying the Markov chain $A \minuso X \minuso Y$ and the distortion constraint
\begin{equation}
\sum_{a,x,y} p_{AXY}(a,x,y) \delta_1(x,\hat{u}) = 0\ ,
\end{equation}
where $\hat{u} = \pi_1(a,y)$. Setting $A = U = \psi_x(X)$ gives $R_{X|Y}^{WZ}(0) \leq H(U|Y)$ and therefore $R_{X|Y}^{WZ}(0) = R_{X|Y}(0) = H(U|Y)$. A similar argument gives $R_{Y|X}(0) = R_{Y|X}^{WZ}(0) = H(V|X)$. The proof is completed by applying Theorem~\ref{Thm:RD-Bounds}.
\end{proof}

Using standard techniques, Theorem~\ref{Thm:RD-Bounds} can be extended from discrete finite alphabets to real-valued alphabets~\cite{Wyner-1978-A-II}. This extension yields the following example for jointly Gaussian sources.

\medskip

\begin{example}[Jointly Gaussian]\label{Exa:RD-Bounds:Exa:1Gaussian}
If $\set{X} = \mathbb{R}_+$, $\set{Y} = \mathbb{R}_+$ and
\begin{align}
\notag
q_{XY}(x,y) =& \frac{1}{2\pi\sigma_x\sigma_y\sqrt{1-\rho^2}} \exp \Bigg[-\frac{1}{2\sigma_x^2\sigma_y^2(1-\rho^2)} \\
\label{Eqn:JointGaussian}
&\qquad \Bigg( \left(\frac{x-m_x}{\sigma_x}\right)^2 + \left(\frac{y-m_y}{\sigma_y}\right)^2
- 2\rho \frac{(x-m_x)(y-m_y)}{\sigma_x \sigma_y}\Bigg)\Bigg]\ ,
\end{align}
and
\begin{subequations}\label{Eqn:Squared-Error}
\begin{align}
\delta_1(x,\hat{x}) &= (x - \hat{x})^2 \\
\delta_2(y,\hat{y}) &= (y - \hat{y})^2\ ,
\end{align}
\end{subequations}
then for all distortion pairs $(d_1,d_2) \in \mathbb{R}_+^2$ we have
\begin{equation}
R(d_1,d_2) = \max\big\{R_{X|Y}(d_1),\ R_{Y|X}(d_2)\big\}\ ,
\end{equation}
where~\cite{Wyner-Jan-1976-A}
\begin{subequations}
\begin{align}
R_{X|Y}(d_1) &= R_{X|Y}^{WZ}(d_1)\ ,\\
R_{Y|X}(d_2) &= R_{Y|X}^{WZ}(d_2)\ ,
\end{align}
\end{subequations}
and
\begin{subequations}\label{Eqn:CRD-Gaussian}
\begin{align}
R_{X|Y}(d_1) &=
\left\{
\begin{array}{ll}
\frac{1}{2}\log \frac{\sigma_x^2(1-\rho^2)}{d_1}, & 0 < d_1 \leq \sigma_x^2(1-\rho^2)\\
0, & \hbox{ otherwise.}
\end{array}
\right.\\
R_{Y|X}(d_2) &=
\left\{
\begin{array}{ll}
\frac{1}{2}\log \frac{\sigma_y^2(1-\rho^2)}{d_2}, & 0 < d_2 \leq \sigma_y^2(1-\rho^2)\\
0, & \hbox{ otherwise.}
\end{array}
\right.
\end{align}
\end{subequations}
\end{example}

\medskip

\begin{remark}
Corollaries~\ref{Thm:RD-Bounds:Cor:Cond-Ind} and~\ref{Thm:RD-Bounds:Cor:2Deterministic} include the results of~\cite[Sec. III.B]{Su-Jun-2010-C} as special cases.  Example~\ref{Exa:RD-Bounds:Exa:1Gaussian} was independently given in~\cite{Su-Jun-2010-C}.
\end{remark}

\medskip

The next result characterises $R(d_1,d_2)$ for one large distortion and shows that the upper bound $R_U(d_1,d_2)$ can be loose. Its proof follows directly from the lower bound $R_L(d_1,d_2)$ in Theorem~\ref{Thm:RD-Bounds} and the coding theorem for the conditional RD function~\cite{Gray-Oct-1972-A}. This proof is omitted.

\medskip

\begin{corollary}\label{Thm:RD-Bounds:Cor:Max-Distortion}
For $(d_1,d_2) \in \reals^2$, we have that
\begin{subequations}
\begin{align}
R(d_1,d_{2,max}) &= R_{X|Y}(d_1)\quad \text{ and}\\
R(d_{1,max},d_2) &= R_{Y|X}(d_2)\ .
\end{align}
\end{subequations}
\end{corollary}

\medskip

In summary, the compress-linear upper bound $R_{U}(d_1,d_2)$ and the cut-set lower bound $R_{L}(d_1,d_2)$ well approximate $R(d_1,d_2)$ when $\delta_1$ and $\delta_2$ are difference distortion measures. Specifically, the ideas of Zamir~\cite{Zamir-Nov-1996-A} can be used to show that the gap between $R_L(d_1,d_2)$ and $R_U(d_1,d_2)$ is no larger than the maximum of two minimax capacities. The bounds yield an exact characterisation of $R(d_1,d_2)$ for sources with zero rate-loss in the Wyner-Ziv problem~\cite{Wyner-Jan-1976-A,Zamir-Nov-1996-A}; however, it is well known that this condition is very restrictive~\cite[Remark 5]{Wyner-Jan-1976-A}. Two sources that satisfy this condition are the jointly Gaussian source with a squared-error distortion measure
(see~\cite[Remark 6]{Wyner-Jan-1976-A} and Example~\ref{Exa:RD-Bounds:Exa:1Gaussian}) and the erasure side-information source with a Hamming distortion measure~\cite{Perron-2007-R,Verdu-Nov-2008-A}. Corollary~\ref{Thm:RD-Bounds:Cor:Max-Distortion} and Example~\ref{Exa:DSBS} demonstrated that the compress-linear upper bound $R_U(d_1,d_2)$  can be loose. We conjecture that $R_L(d_1,d_2)$ is also loose in general, but no counterexample has been found to date. We give a different lower bound for $R(d_1,d_2)$ in Appendix~\ref{App:NewLowerBound}.

\subsection{Kimura-Uyematsu and Heegard-Berger Upper Bounds for $R(d_1,d_2)$}

In this section, we review an upper bound for $R(d_1,d_2)$ that was proposed by Kimura and Uyematsu in~\cite[Thm. 1]{Kimura-Apr-2008-A}, and we compare this bound to the compress-linear upper bound $R_U(d_1,d_2)$. We then formulate a new upper bound for $R(d_1,d_2)$ using a result of Heegard and Berger~\cite{Heegard-Nov-1985-A,Timo-Jun-2010-C}. The main purpose of this section is to unify the achievability results of~\cite{Kimura-Apr-2008-A,Su-Jun-2010-C,Heegard-Nov-1985-A,Timo-Jun-2010-C}.

Let $\set{C}$ be a finite set of cardinality
\begin{equation}
|\set{C}| \leq |\set{X}|\ |\set{Y}| + 2\ .
\end{equation}
Let $\set{P}_{C|XY}(d_1,d_2)$ denote the set channels $p_{C|XY}$ randomly mapping $\set{X}\times\set{Y}$ to $\set{C}$ such that there exist functions $\pi_1 : \set{C} \times \set{Y} \rightarrow \hat{\set{X}}$ and $\pi_2 : \set{C} \times \set{X} \rightarrow \hat{\set{Y}}$ with
\begin{subequations}
\begin{align}
\sum_{x,y,c} p_{C|XY}(c|x,y)q_{XY}(x,y) \delta_1\big(x,\pi_1(c,y)\big) &\leq d_1 \text{ and}\\
\sum_{x,y,c} p_{C|XY}(c|x,y)q_{XY}(x,y) \delta_2\big(y,\pi_2(c,x)\big) &\leq d_2\ .
\end{align}
\end{subequations}
Define
\begin{equation}
R_{U}^*(d_1,d_2) \triangleq \min_{p \in \set{P}_{C|XY}(d_1,d_2)} \max\big\{I(X;C|Y), I(Y;C|X)\big\}\ .
\end{equation}
\medskip
\begin{lemma}[Thm. 1, \cite{Kimura-Apr-2008-A}]\label{Lem:One-Description-Upper}
For $(d_1,d_2) \in \reals^2$, we have that
\begin{equation}
R(d_1,d_2) \leq R_U^*(d_1,d_2)\ .
\end{equation}
\end{lemma}

\medskip

Lemma~\ref{Lem:One-Description-Upper} is called the one-description upper bound because its proof follows from a random coding argument that describes both $\mbf{X}$ and $\mbf{Y}$ with one description.

The one-description bound $R_U^*(d_1,d_2)$ and the compress-linear bound $R_U(d_1,d_2)$ both involve difficult minimizations, so it is not immediately clear when one bound outperforms the other. The next result resolves this question and shows that $R_U^*(d_1,d_2)$ is always better than $R_U(d_1,d_2)$.

\medskip

\begin{lemma}\label{Lem:One-Des-V-Com-Lin}
For $(d_1,d_2) \in \reals^2$, we have that
\begin{equation}
R(d_1,d_2) \leq R_U^*(d_1,d_2) \leq R_U(d_1,d_2)\ .
\end{equation}
\end{lemma}

\medskip

\begin{proof}
We have that
\begin{align}
R_{U}(d_1,d_2) &\equiv \max \Big\{R_{X|Y}^{WZ}(d_1,d_2),\ R_{Y|X}^{WZ}(d_1,d_2)\Big\}\\
&= \max \Bigg\{\min_{p_{AXY}\in\set{P}_{X|Y}^{WZ}(d_1)} I(X;A|Y),\ \min_{p_{BXY}\in\set{P}_{Y|X}^{WZ}(d_2)} I(Y;B|X)\Bigg\}\ ,
\end{align}
where the auxiliary random variables $A$ and $B$ satisfy the Markov chains $A \minuso X \minuso Y$ and $B \minuso Y \minuso X$. Note that $A$ and $B$ do not appear together in any of the mutual information or distortion conditions, so we can combine these minima into a minimum where $A\minuso(X,Y)\minuso B$ forms a Markov chain. To this end, let $\set{P}_{AB|XY}^\ddag(d_1,d_2)$ denote the set of channels $p_{AB|XY}$ mapping $\set{X}$ $\times$ $\set{Y}$ to $\set{A}$ $\times$ $\set{B}$ such that the following properties hold:
\begin{enumerate}
\item The joint distribution, $p_{AB|XY}(a,b|x,y)q_{XY}(x,y)$, factors to form the long Markov chain $A \minuso X \minuso Y \minuso B$.
\item There exist functions $\pi_x:\set{A}\times\set{Y}\rightarrow\Hat{\set{X}}$ and $\pi_y:\set{B}\times\set{X}\rightarrow\Hat{\set{Y}}$ such that
\begin{subequations}
\begin{align}
\sum_{(a,b,x,y)}p_{AB|XY}(a,b|x,y)q_{XY}(x,y)\delta_1\big(x,\pi_x(a,y)\big)&\leq d_1,\\
\sum_{(a,b,x,y)}p_{AB|XY}(a,b|x,y)q_{XY}(x,y)\delta_2\big(y,\pi_y(b,x)\big)&\leq d_2\ .
\end{align}
\end{subequations}
 \end{enumerate}
Note that the long Markov chain $A \minuso X \minuso Y \minuso B$ in condition $1$ is implied by the Markov chains $A \minuso (X,Y) \minuso B$, $A \minuso X \minuso Y$ and $B \minuso Y \minuso X$. We now have that
\begin{align}
\label{Eqn:RD-Coding-RU-2}
R_{U}(d_1,d_2) &= \min_{p_{AB|XY}\in\set{P}_{AB|XY}^{\ddag}(d_1,d_2)} \max \big\{ I(X;A|Y),\ I(Y;B|X)\big\}\ .
\end{align}
The constraint $A \minuso X \minuso Y \minuso B$ implies $(A,X) \minuso Y \minuso B$ which, in turn, implies  $X \minuso (A,Y) \minuso B$. Therefore, we have
\begin{align}
I(X;A|Y) &= H(X|Y) - H(X|A,Y)\\
&= H(X|Y) - H(X|A,B,Y)\\
\label{Eqn:RD-Coding-MC-1}
&= I(X;A,B|Y)\ .
\end{align}
Similarly, we have
\begin{equation}\label{Eqn:RD-Coding-MC-2}
I(Y;B|X) = I(Y;A,B|X)\ .
\end{equation}
Combining~\eqref{Eqn:RD-Coding-RU-2} with~\eqref{Eqn:RD-Coding-MC-1} and~\eqref{Eqn:RD-Coding-MC-2} completes the proof
\begin{align}
R_{U}(d_1,d_2) &= \min_{p_{ABXY}\in\set{P}^{\ddag}(d_1,d_2)} \max \big\{ I(X;A,B|Y),\ I(Y;A,B|X)\big\}\\
\label{Eqn:RD-Coding-RU-4}
&\geq \min_{p_{C|XY}\in\set{P}_{C|XY}(d_1,d_2)} \max \big\{ I(X;C|Y),\ I(Y;C|X)\big\}\ ,
\end{align}
where~\eqref{Eqn:RD-Coding-RU-4} follows because $\set{P}_{C|XY}(d_1,d_2) \supseteq \set{P}_{AB|XY}^{\ddag}(d_1,d_2)$.
\end{proof}

\medskip

The results of Heegard and Berger~\cite[Thm. 2]{Heegard-Nov-1985-A} (see also~\cite{Timo-Jun-2010-C}) can be modified to further strengthen the one-description upper bound. Let $\set{P}_{C|XY}$ denote the set of {\it all} channels $p_{C|XY}$ mapping $\set{X}$ $\times$ $\set{Y}$ to $\set{C}$. For $(d_1,d_2) \in \reals^2$, define
\begin{equation}
R_{U}^{**}(d_1,d_2) \triangleq \min_{p \in \set{P}_{C|XY}} \Big[\max\big\{I(X;C|Y), I(Y;C|X)\big\} + R_{X|CY}(d_1) + R_{Y|CX}(d_2)\Big]\ ,
\end{equation}
where $R_{X|CY}(d_1)$ and $R_{Y|CX}(d_2)$ are the conditional RD functions of $X$ given $(C,Y)$ and $Y$ given $(C,X)$, respectively. The proof of the next result follows directly from~\cite[Thm. 2]{Heegard-Nov-1985-A} and Lemma~\ref{Lem:One-Des-V-Com-Lin} and is omitted.

\medskip

\begin{theorem}\label{Thm:RD*-Upper-Bound}
For $(d_1,d_2) \in \reals^2$, we have that
\begin{equation*}
R(d_1,d_2) \leq R^{**}_{U}(d_1,d_2) \leq R^*_{U}(d_1,d_2) \leq R_{U}(d_1,d_2)\ .
\end{equation*}
\end{theorem}

\medskip

In summary, the compress-linear upper bound $R_U(d_1,d_2)$ and the cut-set lower bound $R_L(d_1,d_2)$ well approximate $R(d_1,d_2)$ for difference distortion measures. The compress-linear bound is weaker than the one-description bound, i.e.  $R_U(d_1,d_2) \geq R_U^*(d_1,d_2)$, and this inequality is strict for the DSBS with Hamming distortion measures (Example~\ref{Exa:DSBS}). Finally, the one-description bound is potentially weaker than Heegard and Berger's bound, i.e. $R_U^*(d_1,d_2) \geq R_U^{**}(d_1,d_2)$; however, we have not found an example where this inequality is strict.  

\subsection{Proof of Theorem~\ref{Thm:CR-RD}}\label{Sec:3:Proof:Thm:CR-RD}

In Theorem~\ref{Thm:CR-RD}, we claimed that the CR-RD function $R_{CR}(d_1,d_2)$ is equal to $R_{CR}^*(d_1,d_2)$. We now prove this result.

\begin{proof}
The coding theorem is a special case of the one-description bound, where $C$ is chosen to be $(\hat{X},\hat{Y})$. We omit the proof. It remains to prove the converse theorem. If $r$ is $(d_1,d_2)$-admissible, then by definition there exists the following:
\begin{enumerate}
\item a monotonically decreasing sequence $\{\epsilon_i\}$ with $\lim_{i\rightarrow\infty} \epsilon_i = 0$, and a monotonically increasing sequence $\{n_i\}$;
\item a sequence of common reconstruction RD codes $\{(f^{(n_i)},$ $g_1^{(n_i)},g_2^{(n_i)},\phi_1^{(n_i)},\phi_2^{(n_i)})\}$, where $\kappa^{(n_i)} \leq r + \epsilon_i$, $\Delta_1^{(n_i)} \leq d_1 + \epsilon_i$, $\Delta_2^{(n_i)} \leq d_2 + \epsilon_i$, $\Pr[\phi_2^{(n_i)}(M,\mbf{X}) \neq g_1^{(n_i)}(M,\mbf{Y})] \leq \epsilon_i$, and $\Pr[\phi_1^{(n_i)}(M,\mbf{Y}) \neq g_2^{(n_i)}(M,\mbf{X})] \leq \epsilon_i$.
\end{enumerate}

We now show that $r+\epsilon_i \geq R^*_{cr}(d_1+\epsilon_i,d_2+\epsilon_i) - \varepsilon(n_i,\epsilon_i)$ for all $i$, where $\lim_{i\rightarrow\infty}\varepsilon(n_i,\epsilon_i) = 0$. To this end, the following inequalities will be useful:
\begin{subequations}\label{Eqn:Thm:CR-RD:Fano1}
\begin{align}
\varepsilon(n_i,\epsilon_i) &\geq H\big(g_2^{(n_i)}(M,\mbf{X})\big|g_1^{(n_i)}(M,\mbf{Y}),\phi_1^{(n_i)}(M,\mbf{Y}),\mbf{Y}\big)\ \text{ and}\\
\varepsilon(n_i,\epsilon_i) &\geq H\big(g_1^{(n_i)}(M,\mbf{Y})\big|g_2^{(n_i)}(M,\mbf{X}),\phi_2^{(n_i)}(M,\mbf{X}),\mbf{X}\big)\ ,
\end{align}
\end{subequations}
where
\begin{equation}
\varepsilon(n_i,\epsilon_i) \triangleq \frac{h(\epsilon_i)}{n_i} + \epsilon_i \log_2 |\hat{\set{X}}\times\hat{\set{Y}}|\ .
\end{equation}
This inequality is a consequence of Fano's inequality~\cite{Cover-2006-B}, the common-reconstruction property
\begin{subequations}
\begin{align}
\Pr[\phi_1^{(n_i)}(M,\mbf{Y}) \neq g_2^{(n_i)}(M,\mbf{X})] &\leq \epsilon_i\ \text{ and}\\
\Pr[\phi_2^{(n_i)}(M,\mbf{X}) \neq g_1^{(n_i)}(M,\mbf{Y})] &\leq \epsilon_i\ ,
\end{align}
\end{subequations}
and the fact that the cardinality of the range of $\phi_i$, $i = 1,2$, can be no more than $|\hat{\set{X}}\times\hat{\set{Y}}|^{n_i}$. Note that $\lim_{i\rightarrow\infty} \varepsilon(n_i,\epsilon_i) = 0$. By definition, we also have
\begin{align}
\label{Eqn:Thm:CR-RD:1:1}
r+\epsilon_i &\geq \kappa^{(n_i)} \equiv \frac{1}{n}\log_2|\set{M}^{(n_i)}|\\
\label{Eqn:Thm:CR-RD:1:2}
&\geq \frac{1}{n_i}H(M) \\
\label{Eqn:Thm:CR-RD:1:3}
&\geq \frac{1}{n_i}H\big(M\big|\mbf{Y}\big) \\
\label{Eqn:Thm:CR-RD:1:5}
&= \frac{1}{n_i}H\big(M,\mbf{Y},g_1^{(n_i)}(M,\mbf{Y}),\phi_1^{(n_i)}(M,\mbf{Y})\big|\mbf{Y}\big)\\
\label{Eqn:Thm:CR-RD:1:6}
&\geq \frac{1}{n_i}H\big(g_1^{(n_i)}(M,\mbf{Y}),\phi_1^{(n_i)}(M,\mbf{Y})\big|\mbf{Y}\big)\\
\notag
&= \frac{1}{n_i}\Big[H\big(g_1^{(n_i)}(M,\mbf{Y}),\phi_1^{(n_i)}(M,\mbf{Y}),g_2^{(n_i)}(M,\mbf{X})\big|\mbf{Y}\big)\\
\label{Eqn:Thm:CR-RD:1:7}
&\qquad\qquad\qquad
-H\big(g_2^{(n)}(M,\mbf{X})\big|g_1^{(n_i)}(M,\mbf{Y}),\phi_1^{(n_i)}(M,\mbf{Y}),\mbf{Y}\big)\Big]\\
\label{Eqn:Thm:CR-RD:1:8}
&\geq \frac{1}{n_i}H\big(g_1^{(n_i)}(M,\mbf{Y}),\phi_1^{(n_i)}(M,\mbf{Y}),g_2^{(n_i)}(M,\mbf{X})\big|\mbf{Y}\big)
- \varepsilon(n_i,\epsilon_i)\\
\label{Eqn:Thm:CR-RD:1:9}
&\geq \frac{1}{n_i} H\big(g_1^{(n_i)}(M,\mbf{Y}),g_2^{(n_i)}(M,\mbf{X})\big|\mbf{Y}\big)
- \varepsilon(n_i,\epsilon_i)\\
\label{Eqn:Thm:CR-RD:1:10}
&\geq \frac{1}{n_i} I\big(\mbf{X};g_1^{(n_i)}(M,\mbf{Y}),g_2^{(n_i)}(M,\mbf{X})\big|\mbf{Y}\big)
- \varepsilon(n_i,\epsilon_i)\\
\label{Eqn:Thm:CR-RD:1:11}
&= \frac{1}{n_i} \sum_{j=1}^{n_i} I\big(X_j;g_1^{(n_i)}(M,\mbf{Y}),g_2^{(n_i)}(M,\mbf{X})\big|\mbf{Y},X_1^{j-1}\big)
- \varepsilon(n_i,\epsilon_i)\\
\label{Eqn:Thm:CR-RD:1:12}
&= \frac{1}{n_i} \sum_{j=1}^{n_i} I\big(X_j;g_1^{(n_i)}(M,\mbf{Y}),g_2^{(n_i)}(M,\mbf{X}),X_1^{j-1},Y_1^{j-1},Y_{j+1}^n\big|Y_i\big)
- \varepsilon(n_i,\epsilon_i)\\
\label{Eqn:Thm:CR-RD:Rate-1}
&\geq \frac{1}{n_i} \sum_{j=1}^{n_i} I\big(X_j;g_1^{(n_i)}(M,\mbf{Y}),g_2^{(n_i)}(M,\mbf{X})\big|Y_j\big)
- \varepsilon(n_i,\epsilon_i)\ ,
\end{align}
where~\eqref{Eqn:Thm:CR-RD:1:1} through~\eqref{Eqn:Thm:CR-RD:1:7} follow from standard identities, \eqref{Eqn:Thm:CR-RD:1:8} follows from~\eqref{Eqn:Thm:CR-RD:Fano1}, \eqref{Eqn:Thm:CR-RD:1:9} through~\eqref{Eqn:Thm:CR-RD:1:11} follow from standard identities, and~\eqref{Eqn:Thm:CR-RD:1:12} follows because the source is iid.

A similar procedure yields
\begin{equation}\label{Eqn:Thm:CR-RD:Rate-2}
r+\epsilon \geq \frac{1}{n_i} \sum_{j=1}^{n_i} I\big(Y_j;g_1^{(n_i)}(M,\mbf{Y}),g_2^{(n_i)}(M,\mbf{X})\big|X_j\big)
- \varepsilon(n_i,\epsilon_i)\ .
\end{equation}

Let $\hat{X}_j$ and $\hat{Y}_j$ denote the $j^{\text{th}}$ elements of $g_1^{(n_i)}(M,\mbf{Y})$ and $g_2^{(n_i)}(M,\mbf{X})$, respectively. I.e. $\hat{X}_j$ and $\hat{Y}_j$ are the $j^{\text{th}}$ symbols reconstructed by the receivers. Expanding the conditions $\Delta_1 \leq d_1 + \epsilon_i$ and $\Delta_2 \leq d_2 + \epsilon_i$ gives
\begin{subequations}
\begin{align}
\label{Eqn:Thm:CR-RD:Dist-1}
\mathbb{E}\left[\frac{1}{n_i}\sum_{j=1}^{n_i} \delta_1(X_j,\hat{X}_j)\right] &\leq d_1 + \epsilon_i\\
\label{Eqn:Thm:CR-RD:Dist-2}
\mathbb{E}\left[\frac{1}{n_i}\sum_{j=1}^{n_i} \delta_2(Y_j,\hspace{0.8mm}\hat{Y}_j)\hspace{0.8mm}\right] &\leq d_2 + \epsilon_i\ .
\end{align}
\end{subequations}
Recall, $\{(X_j,Y_j)\}$ is drawn i.i.d. according to $q_{XY}(x,y)$. For each $j$, let $p_{\hat{X}_j\hat{Y}_j|X_jY_j}(\hat{x}_j,\hat{y}_j|x_j,y_j)$ denote the conditional probability of $(\hat{X}_j,\hat{Y}_j)$ given $(X_j,Y_j)$; that is, combining $p_{\hat{X}_j\hat{Y}_j|X_jY_j}(\hat{x}_j,\hat{y}_j|x_j,y_j)$ with $q_{XY}(x_j,y_j)$ characterises the joint pmf of $(X_j,Y_j,\hat{X}_j,\hat{Y}_j)$. Define the ``time-shared'' channel
\begin{equation}
p_{\hat{X}\hat{Y}|XY}(\hat{x},\hat{y}|x,y) \triangleq \frac{1}{n_i} \sum_{j=1}^n p_{\hat{X}_j\hat{Y}_j|X_jY_j}(\hat{x},\hat{y}|x,y)\ .
\end{equation}
From~\eqref{Eqn:Thm:CR-RD:Dist-1} and \eqref{Eqn:Thm:CR-RD:Dist-2}, we have
\begin{align}
\sum_{x,y,\hat{x},\hat{y}} p_{\hat{X}\hat{Y}|XY}(\hat{x},\hat{y}|x,y) q_{XY}(x,y) &\delta_1(x,\hat{x}) \leq d_1 + \epsilon_i\\
\sum_{x,y,\hat{x},\hat{y}} p_{\hat{X}\hat{Y}|XY}(\hat{x},\hat{y}|x,y) q_{XY}(x,y) &\delta_2(y,\hat{y}) \leq d_2 + \epsilon_i\ ;
\end{align}
consequently, $p_{\hat{X}\hat{Y}|XY}(\hat{x},\hat{y}|x,y) \in \set{P}_{\hat{X}\hat{Y}|XY}(d_1+ \epsilon_i,d_2+ \epsilon_i)$. We have that
\begin{align}
\label{Eqn:Thm:CR-RD:Rate-3}
\frac{1}{n_i} \sum_{j=1}^{n_i} I\big(X_j;\hat{X}_j,\hat{Y}_j\big|Y_j\big) &\geq I(X;\hat{X},\hat{Y}|Y)\ , \text{ and}\\
\label{Eqn:Thm:CR-RD:Rate-4}
\frac{1}{n_i} \sum_{j=1}^{n_i} I\big(Y_j;\hat{X}_j,\hat{Y}_j\big|X_j\big) &\geq I(Y;\hat{X},\hat{Y}|X) \ ,
\end{align}
where we have used Jensen's inequality together with the convexity of $I(X;\hat{X},\hat{Y}|Y)$ and $I(Y;\hat{X},\hat{Y}|X)$ in $p_{\hat{X}\hat{Y}|XY}$ when the joint pmf of $(X,Y)$ (here $q_{XY}$) is fixed (see Lemma~\ref{Lem:CMI-Convex} below). Finally, combining~\eqref{Eqn:Thm:CR-RD:Rate-3} and~\eqref{Eqn:Thm:CR-RD:Rate-4} with the definition of $R_{CR}^*(d_1,d_2)$  we have
\begin{align}
r + \epsilon_i &\geq \max \big\{ I(X;\hat{X},\hat{Y}|Y),\ I(Y;\hat{X},\hat{Y}|X) \big\} - \varepsilon(n_i,\epsilon_i)\\
&\geq R_{CR}^*(d_1 +\epsilon_i,d_2 + \epsilon_i) - \varepsilon(n_i,\epsilon_i) \ ,
\end{align}
which is the desired result.

The converse is completed by noting that $\lim_{i\rightarrow\infty} \epsilon_i = 0$, $\lim_{i\rightarrow\infty}\varepsilon(n_i,\epsilon_i) = 0$, and $R_{CR}^*(d_1,d_2)$ is a continuous function of $d_1$ and $d_2$.
\end{proof}

\medskip

\begin{lemma}\label{Lem:CMI-Convex}
Suppose the random vector $(A,B,C)$ on $\set{A} \times \set{B} \times \set{C}$ is characterised by the joint pmf $p_{ABC}(a,b,c) = p_{C|AB}(c|a,b)p_{AB}(a,b)$. The condition mutual information $I(A;C|B)$ is convex in $p_{C|AB}(c|a,b)$ for fixed $p_{AB}(a,b)$.
\end{lemma}

\begin{proof}
Fix $p_{AB}$. From the convexity of mutual information~\cite[Thm. 2.7.4]{Cover-2006-B}, we have that $I(A;C|B=b)$ is convex in $p_{C|AB}(\cdot|\cdot,b)$ for each $b$. The lemma follows by noting that $I(A;C|B)$ is a convex combination of $I(A;C|B=b)$. Further details can be found in Appendix~\ref{App:Lemma-Convexity}.
\end{proof}

\subsection{Extreme Distortions}

The next result shows that if one source is required to be reconstructed with vanishing Hamming distortion, then the RD function $R(d_1,d_2)$ and the CR-RD function $R_{CR}(d_1,d_2)$ both collapse to the cut-set lower bound $R_L(d_1,d_2)$.

\medskip

\begin{corollary}\label{Thm:CR-RD:Cor:OneLossless}
If $\delta_1$ is a Hamming distortion measure, then for all $d_2 \in \reals$ we have that
\begin{equation}
R(0,d_2) = R_{CR}(0,d_2) = \max\big\{H(X|Y),R_{Y|X}(d_2)\big\}\ .
\end{equation}
\end{corollary}

\medskip

\begin{proof}
From Proposition~\ref{Pro:RC-CR-Convex} and Theorems~\ref{Thm:CR-RD} and~\ref{Thm:RD-Bounds} we have that
\begin{equation}\label{Eqn:Proof-Extreme1}
R_L(0,d_2) \leq R(0,d_2) \leq R_{CR}(0,d_2) = \min_{p_{\hat{X}\hat{Y}|XY}\in\set{P}_{\hat{X}\hat{Y}|XY}(0,d_2)} \max\big\{I(X;\hat{X},\hat{Y}|Y),I(Y;\hat{X},\hat{Y}|X)\big\}\ .
\end{equation}
Let $p_{\hat{Y}|XY}$ be a channel that achieves the minimum for the conditional RD function $R_{Y|X}(d_2)$. This channel and $q_{XY}$ together define a joint pmf for $(X,Y,\hat{Y})$. In addition, set $\hat{X} = X$ to obtain a joint pmf for $(X,Y,\hat{X},\hat{Y})$. This joint pmf belongs to the set $\set{P}_{\hat{X}\hat{Y}|XY}(0,d_2)$. Note, we have the Markov chain $(Y,\hat{Y}) \minuso X \minuso \hat{X}$ and therefore the chain $Y \minuso (X,\hat{Y}) \minuso \hat{X}$. On substituting this joint pmf into~\eqref{Eqn:Proof-Extreme1} we obtain the following upper bound for $R_{CR}(0,d_2)$:
\begin{align}
R_{CR}(0,d_2) &\leq \max\big\{ I(X;\hat{X},\hat{Y}|Y),I(Y;\hat{X},\hat{Y}|X)\big\}\\
&= \max\{H(X|Y)-H(X|Y,\hat{X},\hat{Y}),I(Y;\hat{Y}|X)+I(Y;\hat{X}|X,\hat{Y})\big\}\\
\label{Eqn:Cor:OneLosslessRecon-1}
&= \max\{H(X|Y),I(Y;\hat{Y}|X)\}\\
\label{Eqn:Cor:OneLosslessRecon-2}
&= \max\{H(X|Y),R_{Y|X}(d_2)\big\}\ ,
\end{align}
where~\eqref{Eqn:Cor:OneLosslessRecon-1} follows because $\hat{X} = X$ and $Y \minuso (X,\hat{Y}) \minuso \hat{X}$ forms a Markov chain, and~\eqref{Eqn:Cor:OneLosslessRecon-2} follows because $p_{\hat{Y}|X,Y}$ was chosen to achieve the minimum in the definition of $R_{Y|X}(d_2)$. The proof is completed by noting that
\begin{align}
R_L(0,d_2) &\triangleq \max\big\{R_{X|Y}(0),R_{Y|X}(d_2)\big\}\\
&= \max\big\{H(X|Y),R_{Y|X}(d_2)\big\}\ .
\end{align}
\end{proof}

\medskip

The next result covers the one large distortion setting. The proof follows directly from Theorem~\ref{Thm:CR-RD} and is omitted. Note that it may differ from Corollary~\ref{Thm:RD-Bounds:Cor:Max-Distortion} -- the large distortion result for $R(d_1,d_2)$.

\medskip

\begin{corollary}\label{Thm:CR-RD:Cor:Max-Distortion}
For $d_1 \in \reals$ we have that
\begin{equation}
R_{CR}(d_1,d_{2,max}) = \min_{\set{P}_{\hat{X}|XY}(d_1)} \max \big\{ I(X;\hat{X}|Y), I(Y;\hat{X}|X)\big\}\ ,
\end{equation}
where $\set{P}_{\hat{X}|XY}$ denotes the set of all test channels $p_{\hat{X}|XY}$ mapping $\set{X} \times \set{Y}$ to $\hat{\set{X}}$ such that
\begin{equation}
\sum_{x,y,\hat{x}} p_{\hat{X}|XY}(\hat{x}|x,y)q_{XY}(x,y)\delta_1(x,\hat{x}) \leq d_1\ .
\end{equation}
\end{corollary}

\medskip

\subsection{Small Distortions and a Proof of Theorem~\ref{Thm:Small-Distortions}}\label{Sec:3:Small-Distortions}

The following result gives a useful upper bound for $R_{CR}(d_1,d_2)$. We will use this bound to prove the small distortion result Theorem~\ref{Thm:Small-Distortions}.

\medskip

\begin{corollary}\label{Cor:Thm:CR-RD}
For $(d_1,d_2) \in \reals^2$ we have that
\begin{align}
R_{CR}(d_1,d_2) \leq \max\big\{R_{XY}(d_1,d_2) - R_X(d_1), R_{XY}(d_1,d_2) - R_Y(d_2) \big\}\ .
\end{align}
\end{corollary}

\medskip

\begin{proof}
Let $p_{\hat{X}\hat{Y}|XY}$ achieve the minimum for the joint rate distortion function $R_{XY}(d_1,d_2)$. Then,
\begin{equation}
R_{CR}(d_1,d_2) \leq R_{XY}(d_1,d_2) - \min \Big\{I(X;\hat{X},\hat{Y}),\ I(Y;\hat{X},\hat{Y})\Big\}\ ,
\end{equation}
where the remaining mutual information terms are evaluated using $p_{\hat{X}\hat{Y}|XY}\ \cdot\ q_{XY}$. Note that
\begin{equation}
I(X;\hat{X},\hat{Y}) \geq I(X;\hat{X}) \geq R_X(d_1)\ ,
\end{equation}
where the last inequality follows from the definition of $R_X(d_1)$. Similarly, we also have that $I(Y;\hat{X},\hat{Y}) \geq R_Y(d_2)$, and thus
\begin{equation}
R_{CR}(d_1,d_2) \leq R_{XY}(d_1,d_2) - \min \Big\{ R_X(d_1),\ R_y(d_2) \Big\}\ .
\end{equation}
\end{proof}

On combining this result with Proposition~\ref{Pro:RC-CR-Convex} and Theorem~\ref{Thm:RD-Bounds}, we have
\begin{align}
\max\big\{ R_{XY}(d_1,d_2) - R_X(d_1), R_{XY}(d_1,d_2) - R_Y(d_2) \big\} &\geq R_{CR}(d_1,d_2)\\
&\geq R(d_1,d_2)\\
\label{Eqn:Cor:Thm:CR-RD-3}
&\geq \max\big\{R_{X|Y}(d_1), R_{Y|X}(d_2)\big\}\ .
\end{align}
From this chain of inequalities, it is clear that if
\begin{subequations}\label{Eqn:ImTired1}
\begin{align}
R_{X|Y}(d_1) &= R_{XY}(d_1,d_2) - R_{Y}(d_2)\quad\text{ and}\\
R_{Y|X}(d_2) &= R_{XY}(d_1,d_2) - R_{X}(d_1)\ ,
\end{align}
\end{subequations}
then we have that the RD function $R(d_1,d_2)$ and the CR-RD function $R_{CR}(d_1,d_2)$ both meet the cut-set lower bound  $R_L(d_1,d_2)$. The next two examples give situations where~\eqref{Eqn:ImTired1} holds.

\medskip

\begin{example}
If $X$ and $Y$ are independent ($q_{XY} = q_X \cdot q_Y$), then
\begin{equation}
\max\{R_{X|Y}(d_2),\ R_{Y|X}(d_2)\} = \max\{R_X(d_1), R_Y(d_2)\}\ ,
\end{equation}
and
\begin{align}
R_{XY}(d_1,d_2) &- \min \big\{ R_X(d_1),R_Y(d_2) \big\} \\
&= R_X(d_1) + R_Y(d_2) - \min \big\{ R_X(d_1),R_Y(d_2) \big\}\\
&= \max \big\{ R_X(d_1),R_Y(d_2) \big\}\ ;
\end{align}
therefore,
\begin{align}
R(d_1,d_2) = R_{CR}(d_1,d_2) = \max \{R_X(d_1),\ R_Y(d_2)\}\ .
\end{align}
\end{example}
\medskip
\begin{example}
If $\delta_1$ and $\delta_2$ are Hamming measures, then
\begin{align}
\max\{R_{X|Y}(0),\ R_{Y|X}(0)\} = \max\{H(X|Y),\ H(Y|X)\}
\end{align}
and
\begin{align}
R_{XY}(d_1,d_2) &- \min \big\{ R_X(d_1), R_Y(d_2) \big\} \\
&= H(X,Y) - \min \big\{ H(X), H(Y) \big\}\\
&= \max \big\{H(X|Y), H(Y|X)\big\}\ ;
\end{align}
therefore,
\begin{align}
R(0,0) = R_{CR}(0,0) = \max \{H(X|Y),\ H(Y|X)\}\ .
\end{align}
\end{example}

\medskip

This idea of matching the lower and upper bounds in~\eqref{Eqn:Cor:Thm:CR-RD-3} is not just useful for these simple examples. Our main result, Theorem~\ref{Thm:Small-Distortions}, showed that it is also useful for sources with Hamming distortions with small distortions. The proof of this result is a simple consequence of Corollary~\ref{Cor:Thm:CR-RD}.

\begin{proof}[Proof of Theorem~\ref{Thm:Small-Distortions}]
Let us recall Gray's results for the extended Shannon lower bounds of joint, conditional and marginal RD functions. Specifically, from~\cite[Thm. 3.2 $\&$ Cor. 3.2]{Gray-Jul-1973-A} there exists a strictly positive surface $\set{D}$ in $\mathbb{R}_+^2$ such that
\begin{subequations}
\begin{align}
R_{XY}(d_1,d_2) &= R_{X|Y}(d_1) + R_{Y}(d_2)\ , \text{ and}\\
R_{XY}(d_1,d_2) &= R_{Y|X}(d_2) + R_{X}(d_1)\
\end{align}
\end{subequations}
for all $(d_1,d_2) \in \mathbb{R}_+^2$ that lies on or below $\set{D}$. Combining this result with~\eqref{Eqn:Cor:Thm:CR-RD-3} proves the theorem.
\end{proof}

\subsection{Proof of Theorem~\ref{Thm:DSBS-CR}}\label{Sec:3:DSBS}

\begin{figure}[t]
\begin{center}
\includegraphics[width=85mm]{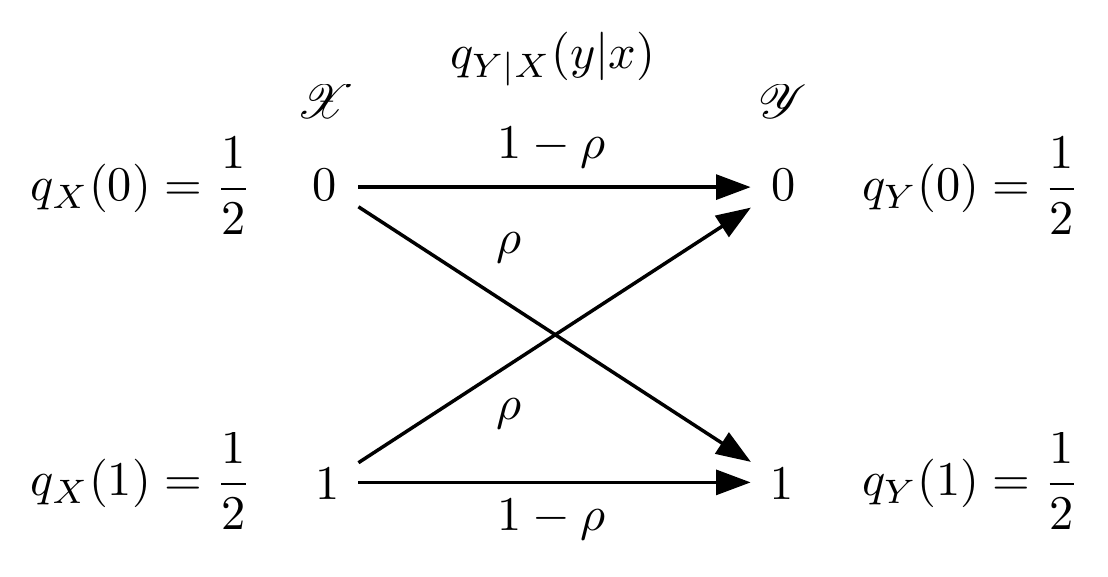}\\
\caption{Doubly Symmetric Binary Source (DSBS) with cross over probability $\rho$.}\label{Fig:DSBS-PMF}
\end{center}
\end{figure}

The joint pmf $q_{XY}$ of the DSBS can be thought of as resulting from using $X$ as a uniform input to a binary symmetric channel (BSC) with cross over probability $\rho$, see Figure~\ref{Fig:DSBS-PMF}. By symmetry, we can also think of $q_{XY}$ resulting from using $Y$ as a uniform input to a BSC with cross over probability $\rho$.

\medskip

\subsubsection{Proof of~\eqref{Thm:DSBS-CR:Part1}}

In Example~\ref{Exa:DSBS}, it was shown that the RD function without common reconstructions $R(d_1,d_2)$ equals the cut-set lower bound $R_L(d_1,d_2)$. Since $R_{CR}(d,d) \geq R(d,d)$ for all $d \in [0,1/2]$, we have that
\begin{equation}
R_{CR}(d,d) \geq R(d,d) =
\left\{
  \begin{array}{ll}
    h(\rho) - h(d), & \hbox{ for } d \leq \rho \\
    0, & \hbox{ for } d > \rho\ .
  \end{array}
\right.
\end{equation}
Let
\begin{equation}
d^* = \frac{1}{2} - \frac{1}{2} \sqrt{1-2\rho}\
\end{equation}
and note that $d^* \leq \rho$. For any $d \in [0,d^*]$, we now construct a test channel $p_{\hat{X}\hat{Y}|XY}$ that belongs to $\set{P}_{\hat{X}\hat{Y}|XY}(d,d)$ and $I(X;\hat{X},\hat{Y}|Y) = I(Y;\hat{X},\hat{Y}|X) = h(\rho) - h(d)$.

\begin{figure}[t]
\begin{center}
\includegraphics[width=120mm]{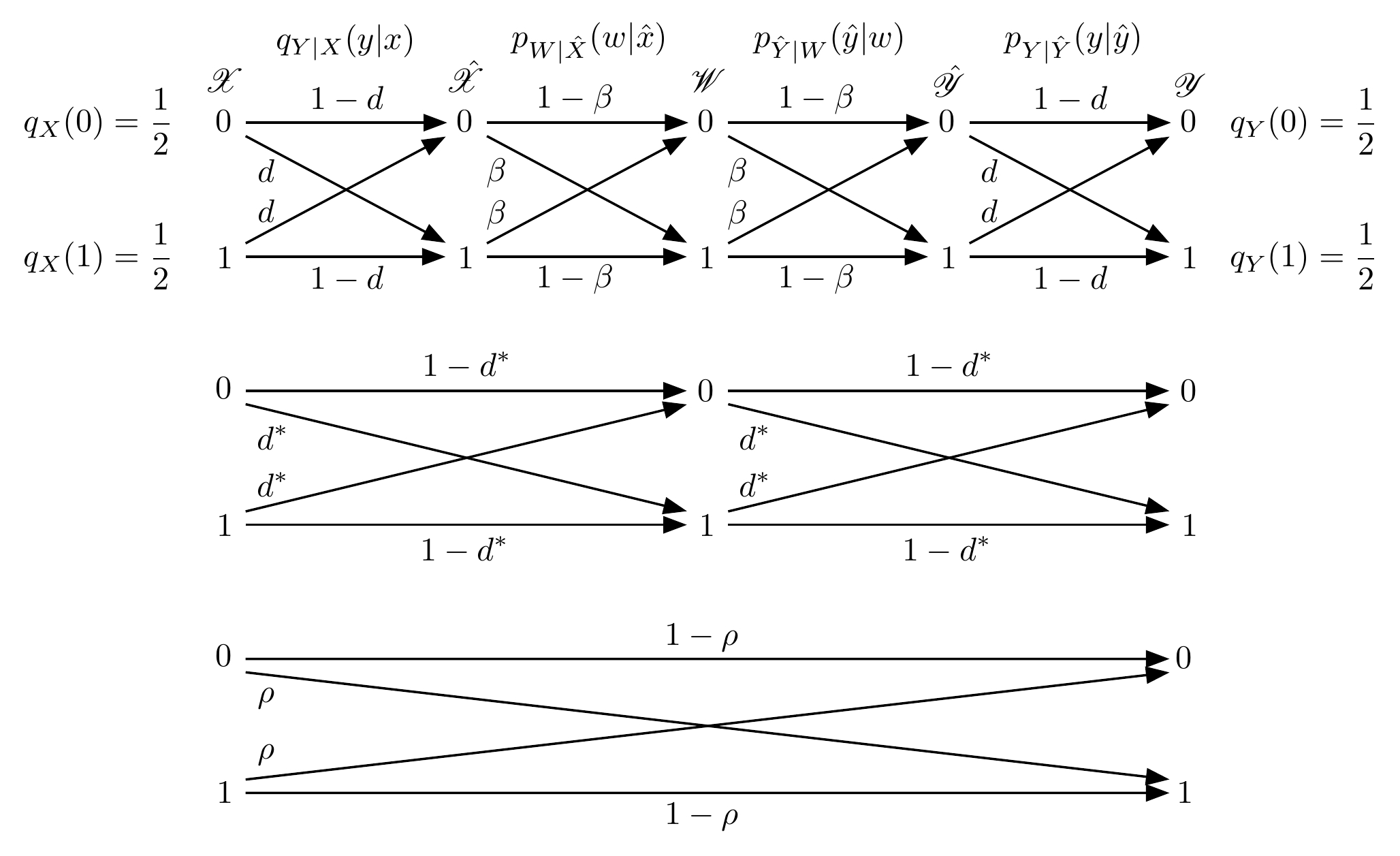}\\
\caption{DSBS test channel configuration for $d < d^*$.}\label{Fig:DSBS-Test-Channel}
\end{center}
\end{figure}

Fix $d \in [0,d^*]$, and let
\begin{equation}
\beta = \frac{d^*-d}{1-2d}\ .
\end{equation}
Note that $d \star \beta = d^*$, where $d \star \beta \triangleq d(1-\beta) + (1-d)\beta$ is the binary convolution. Let $\set{W} = \{0,1\}$. We now define a joint pmf $p(x,\hat{x},w,\hat{y},y)$ on $\set{X} \times \hat{\set{X}} \times \set{W} \times \hat{\set{Y}} \times \set{Y}$ by assuming a uniform input to the cascade of the four BSCs shown in Figure~\ref{Fig:DSBS-Test-Channel}. Specifically, we set
\begin{equation}\label{Eqn:CR:DSBS:Channel1}
p(x,\hat{x},w,\hat{y},y) = p(x)p(\hat{x}|x)p(w|\hat{x})p(\hat{y}|w)p(y|\hat{y})\ ,
\end{equation}
where $p(x) = 1/2$ for $x = 0$ and $x = 1$ and
\begin{subequations}
\begin{align}
p(\hat{x}|x) &= (1-d) \mbf{1}_{\hat{x},x} + d (1-\mbf{1}_{\hat{x},x})\\
p(w|\hat{x}) &= (1-\beta) \mbf{1}_{w,\hat{x}} + \beta (1-\mbf{1}_{w,\hat{x}})\\
p(\hat{y}|w) &= (1-\beta) \mbf{1}_{\hat{y},w} + \beta (1-\mbf{1}_{\hat{y},w})\\
p(y|\hat{y}) &= (1-d) \mbf{1}_{y,\hat{y}} + \beta (1-\mbf{1}_{y,\hat{y}})\ .
\end{align}
\end{subequations}
Note that since $p(x)$ is uniform we may equivalently view $p(x,\hat{x},w,\hat{y},y)$ as resulting from using $p(y)$ as a uniform input to the (reverse) cascade of four BSCs shown in Figure~\ref{Fig:DSBS-Test-Channel}.

By construction, the expected distortions $\mathbb{E}[\delta_1(X,\hat{X})]$ and $\mathbb{E}[\delta_2(Y,\hat{Y})]$ for this joint pmf are both equal to $d$. Moreover, since $d \star \beta = d^*$ and $d^* \star d^* = \rho$ we have that
\begin{equation}
\sum_{\hat{x},w,\hat{y}} p(x,\hat{x},w,\hat{y},y) = q_{XY}(x,y)\ ,
\end{equation}
and the joint pmf $p(x,\hat{x},w,\hat{y},y)$ defines a valid channel in $\set{P}_{\hat{X}\hat{Y}|XY}(d,d)$. Combining this channel with Theorem~\ref{Thm:CR-RD} yields
\begin{align}
R_{CR}(d,d) &\leq \max \big\{I(X;\hat{X},\hat{Y}|Y),I(Y;\hat{X},\hat{Y}|X)\big\}\\
&= \max\big\{ H(X|Y) - H(X|Y,\hat{X},\hat{Y}), H(Y|X) - H(Y|X,\hat{X},\hat{Y})\big\}\\
\label{Eqn:Thm:DSBS-CR:1}
&= \max\big\{ H(X|Y) - H(X|\hat{X}),H(Y|X)-H(Y|\hat{Y})\big\}\\
\label{Eqn:Thm:DSBS-CR:2}
&= h(\rho) - h(d)\ ,
\end{align}
where~\eqref{Eqn:Thm:DSBS-CR:1} follows because $X \minuso \hat{X} \minuso (Y,\hat{Y})$ and $Y \minuso \hat{Y} \minuso (X,\hat{X})$ form a Markov chains, and~\eqref{Eqn:Thm:DSBS-CR:2} follows by construction.

\subsubsection{Proof of~\eqref{Thm:DSBS-CR:Part2}}

At $d_1 = d_2 = d^*$, we have that
\begin{equation}
R(d^*,d^*) = R_{CR}(d^*,d^*) = h(\rho) - h(d^*)\ .
\end{equation}
The marginal RD functions of $X$ and $Y$ are given by
\begin{subequations}
\begin{align}
R_X(d^*) &= 1 - h(d^*)\quad \text{ and}\\
R_Y(d^*) &= 1 - h(d^*)\ .
\end{align}
\end{subequations}
Wyner showed that the common information of $X$ and $Y$ is given by~\cite[Eqn. 1.19]{Wyner-Mar-1975-A}
\begin{equation}
K(X;Y) = 1 + h(\rho) - 2h(d^*)\ .
\end{equation}
Therefore, $R_{CR}(d^*,d^*) = K(X;Y) - R_X(d^*)$.

\medskip

\begin{remark}
The $W$ that achieves the minimum for $K(X;Y)$ is the same as the $W$ in Figure~\ref{Fig:DSBS-Test-Channel}. Specifically, for $d_1 = d_2 = d^*$ we use $\hat{X} = \hat{Y} = W$.
\end{remark}

\medskip

\subsubsection{Proof of~\eqref{Thm:DSBS-CR:Part3a}}

Suppose that $d_1 = d_2 = d$. If $d = 0$, then it is clear that $R(0,0) = R_{CR}(0,0) = H(X|Y) = H(Y|X) = h(\rho)$. Moreover, it is optimal to choose \begin{equation}
p_{\hat{X}\hat{Y}|XY}(\hat{x},\hat{y}|x,y) = \left\{
                                               \begin{array}{ll}
                                                 1, & \hbox{ if } x = \hat{x} \text{ and } y = \hat{y}\\
                                                 0, & \hbox{ if } x \neq \hat{x} \text{ or } y \neq \hat{y}\ .
                                               \end{array}
                                             \right.
\end{equation}
With this choice of test channel, we have that $X \minuso (\hat{X},\hat{Y}) \minuso Y$ forms a Markov chain. The next lemma shows that this chain is necessary for $R_{CR}(d,d) = R(d,d)$.

\bigskip

\begin{lemma}\label{Lem:DSBS-Contradiction}
If $R_{CR}(d,d) = R(d,d)$, then the minimum
\begin{equation}\label{Eqn:CR-MC}
\min_{p_{\hat{X}\hat{Y}|XY} \in \set{P}_{\hat{X}\hat{Y}|XY}(d,d)} \max\Big\{I(X;\hat{X},\hat{Y}|Y),\ I(Y;\hat{X},\hat{Y}|X)\Big\}
\end{equation}
is achieved by a test-channel $p^*_{\hat{X}\hat{Y}|XY}$ for which the resultant joint pmf for $(X,Y,\hat{X},\hat{Y})$ factors to form the Markov chain $X \minuso (\hat{X},\hat{Y}) \minuso Y$ and $H(X|\hat{X},\hat{Y}) = H(Y|\hat{X},\hat{Y}) = h(d)$.
\end{lemma}

\bigskip

\begin{proof}
Suppose $p^*_{\hat{X}\hat{Y}|XY}$ achieves the minimum in~\eqref{Eqn:CR-MC}. From the definition of conditional rate-distortion function, $R_{X|Y}(d)$, the following is apparent
\begin{equation}\label{Eqn:CR-CI-1}
I(X;\hat{X},\hat{Y}|Y) \geq I(X;\hat{X}|Y)
\geq R_{X|Y}(d) = h(\rho) - h(d)\ .
\end{equation}
Similarly,
\begin{equation}\label{Eqn:CR-CI-2}
I(Y;\hat{X},\hat{Y}|X) \geq I(Y;\hat{Y}|X) \geq R_{Y|X}(d) = h(\rho) - h(d)\ .
\end{equation}
If $\max\{I(X;\hat{X},\hat{Y}|Y),\ I(Y;\hat{X},\hat{Y}|X)\} = h(\rho) - h(d)$
then from~\eqref{Eqn:CR-CI-1} and~\eqref{Eqn:CR-CI-2} and $H(X|Y) = H(Y|X) = h(\rho)$, we have
\begin{equation}
\label{Eqn:CR-CI-3}
H(X|\hat{X},\hat{Y},Y) = H(Y|\hat{X},\hat{Y},X) = h(d)\ .
\end{equation}
Then, we further have
\begin{align}
I(X;Y|\hat{X},\hat{Y}) &= H(X|\hat{X},\hat{Y}) - H(X|\hat{X},\hat{Y},Y)\\
&= H(X|\hat{X},\hat{Y}) - h(d)\\
&= H(X\oplus\hat{X}|\hat{X},\hat{Y}) - h(d)\\
&\leq H(X\oplus\hat{X}) - h(d)\\
&\leq 0\ .
\end{align}
The non-negativity of conditional mutual information gives $I(X;Y|\hat{X},\hat{Y}) = 0$ and therefore $X \minuso (\hat{X},\hat{Y}) \minuso Y$. The proof is completed by combining this chain with~\eqref{Eqn:CR-CI-3} to get $H(X|\hat{X},\hat{Y}) = H(Y|\hat{X},\hat{Y}) = h(d)$.
\end{proof}

\bigskip

The proof of~\eqref{Thm:DSBS-CR:Part3a} will follow via a contradiction. Suppose there exists $d > d^*$ such that $R_{CR}(d,d) = h(\rho) - h(d)$. From Theorem~\ref{Thm:DSBS-CR} we have that
\begin{align}\label{Eqn:DSBS:Contradiction-1}
R_{CR}(d,d) &= \min_{p_{\hat{X}\hat{Y}|XY}\in\set{P}_{\hat{X}\hat{Y}}(d,d)} \Big[I(X,Y;\hat{X},\hat{Y}) - \min \big\{I(X;\hat{X},\hat{Y}),I(Y;\hat{X},\hat{Y})\big\}\Big]\ .
\end{align}
Let $p_{\hat{X}\hat{Y}|XY}$ be the test channel that achieves the indicated minimum, and consider the term $I(X,Y;\hat{X},\hat{Y})$ in~\eqref{Eqn:DSBS:Contradiction-1}. From Lemma~\ref{Lem:DSBS-Contradiction}, the joint pmf induced by $p_{\hat{X}\hat{Y}|XY}$ and $q_{XY}$ factors to form the Markov chain $X \minuso (\hat{X},\hat{Y}) \minuso Y$; therefore, $I(X,Y;\hat{X},\hat{Y})$ can be lower bounded by Wyner's common information~\cite[Sec. 3]{Wyner-Mar-1975-A} via
\begin{equation}
I(X,Y;\hat{X},\hat{Y}) \geq 1 + h(\rho) - 2h(d^*)\ .
\end{equation}
We have $H(Y) = 1$, and from Lemma~\ref{Lem:DSBS-Contradiction} we have $H(X|\hat{X},\hat{Y}) = H(Y|\hat{X},\hat{Y}) = h(d)$. Since $h(\rho) - h(d)$ is strictly decreasing on $[0,\rho)$ it follows that $R_{CR}(d,d) < R_{CR}(d^*,d^*)$, which is equivalent to
\begin{equation}
h(\rho) - h(d^*) > I(X,Y;\hat{X},\hat{Y}) - I(Y;\hat{X},\hat{Y})\ ,
\end{equation}
and by the above discussion
\begin{equation}
h(\rho) - h(d) > 1 + h(\rho) - 2h(d^*) - \big[1 - h(d)\big]\ ,
\end{equation}
which implies $h(d^*) > h(d)$, which is a contradiction since $h(\cdot)$ is strictly increasing on $[0,1/2]$.

\subsubsection{Proof of~\eqref{Thm:DSBS-CR:Part3b}}

\begin{figure}
\begin{center}
  \includegraphics[width=90mm]{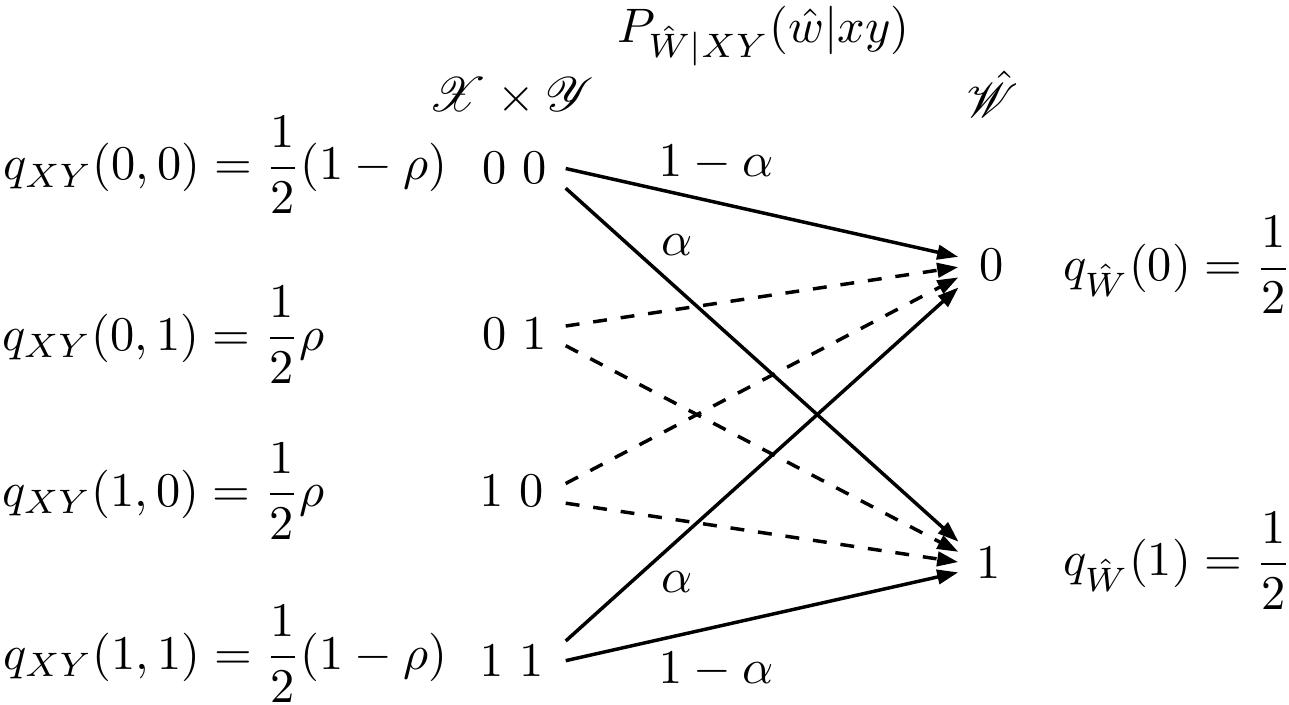}\\
\end{center}
\caption{Depiction of the channel $p_{\hat{W}|XY}(\hat{w}|x,y)$. The transitions represented by dotted lines each have probability $p_{\hat{W}|XY}(\hat{w}|x,y) = 1/2$.}\label{Fig:FourInputChannel}
\end{figure}

We choose a channel $p_{\hat{X}\hat{Y}|XY}$ that achieves the bound given in~\eqref{Thm:DSBS-CR:Part3b}. Let $\hat{\set{W}} \triangleq \{0,1\}$, and define
\begin{equation}
p_{\hat{W}|XY}(\hat{w}|x,y) \triangleq (1-\alpha)(1-\mbf{1}_{x,y})(1-\mbf{1}_{x,y,\hat{w}}) + \alpha (1-\mbf{1}_{x,y}) \mbf{1}_{x,y,\hat{w}} + \frac{1}{2} \mbf{1}_{x,y}\ ,
\end{equation}
where
\begin{equation}
\alpha \triangleq \frac{2d-\rho}{2(1-\rho)}\ ,
\end{equation}
and $\mbf{1}_{x,y}$ and $\mbf{1}_{x,y,z}$ are indicator functions (equal one if the subscripts are equal and zero otherwise). The channel $p_{\hat{W}|XY}(\hat{w}|x,y)$ is depicted in Figure~\ref{Fig:FourInputChannel}.

Set $\hat{X} = \hat{W}$ and $\hat{Y} = \hat{W}$. Note that
\begin{subequations}
\begin{align}
p_{\hat{W}|X}(\hat{w}|x) &\triangleq \sum_{y \in \set{Y}} p_{\hat{W}|XY}(\hat{w}|x,y) q_{Y|X}(y|x)\text{ and}\\
p_{\hat{W}|Y}(\hat{w}|y) &\triangleq \sum_{x \in \set{X}} p_{\hat{W}|XY}(\hat{w}|x,y) q_{X|Y}(x|y)\ ,
\end{align}
\end{subequations}
are both BSCs with a crossover probability $d$. Therefore, $\mathbb{E}[\delta_1(X,\hat{X})] = d$ and $\mathbb{E}[\delta_2(Y,\hat{Y})]=d$. Finally, the rate of the channel is given by
\begin{align}
I(X;\hat{W}|Y) &= H(\hat{W}|Y) - H(\hat{W}|X,Y)\\
&= h(d) - \big[\rho + (1-\rho) h(\alpha)\big]\ .
\end{align}
By symmetry, we also have $I(X;\hat{W}|Y) = h(d) - \rho - (1-\rho) h(\alpha)$, which completes the proof.

\medskip

\begin{remark}
The channel $p_{\hat{W}|XY}(\hat{w}|x,y)$ can be view as the natural continuation of the channel~\eqref{Eqn:CR:DSBS:Channel1}, which was used to prove~\eqref{Thm:DSBS-CR:Part1}. Specifically, $p_{\hat{W}|XY}(\hat{w}|x,y)$ is formed by passing $W$ through a BSC with crossover probability $(d-d^*)/(1-2d^*)$. This latter quantity is chosen because
\begin{equation}
d = d^* \star \frac{d-d^*}{1-2d^*}\ .
\end{equation}
\end{remark}


\section{Joint Source-Channel Coding: Auxiliary Results $\&$ Proofs}\label{Sec:5}

We now extend the source coding results of Section~\ref{Sec:4} to the joint source-channel coding setting (Definitions~\ref{Def:JSCC} and~\ref{Def:JSCC-CR}). We begin by proving Theorem~\ref{Thm:JSCC-Cut-Set}. 

\subsection{Proof of Theorem~\ref{Thm:JSCC-Cut-Set}}

\begin{proof}
If $(d_1,d_2) \in \mathbb{R}_+^2$ is admissible with bandwidth expansion factor $\kappa$, then by definition there exists for every $\epsilon > 0$ a joint source-channel code $(f^{(t)},g_1^{(t)},g_2^{(t)})$ with $\Delta_i^{(\kappa_s t)} \leq d_i + \epsilon$, $i = 1,2$.

Let $\mbf{W} = W_1,W_2,\ldots,W_{\kappa_c t}$ denote the codeword that is produced by the encoder. Let $p_{W_i}$ denote the marginal pmf for the $i^{\text{th}}$ symbol $W_i$.  Define a new ``time-shared'' random variable $\tilde{W}$ on $\set{W}$ with pmf
\begin{equation}
p_{\tilde{W}}(w) \triangleq \frac{1}{\kappa_c t} \sum_{i=1}^{\kappa_c t} p_{W_i}(w)\ .
\end{equation}

Since $I(\tilde{W};U)$ is a concave function for fixed $Q_{U|\tilde{W}}$, we have from Jensen's inequality
\begin{equation}\label{Eqn:Tuncel-Convexity}
I(\tilde{W};U) \geq \frac{1}{\kappa_c t}\sum_{i=1}^{\kappa_c t} I(W_i;U_i)\ .
\end{equation}
We further have
\begin{align}
I(\mbf{W};\mbf{U}) &= H(\mbf{U}) - H(\mbf{U}|\mbf{W})\\
&= \sum_{i=1}^{\kappa_c t} \Big[ H(U_i|U_1,U_2,\ldots,U_{i-1}) - H(U_i|\mbf{W},U_1,U_2,\ldots,U_{i-1})  \Big]\\
\label{Eqn:BC:Gerhard-1}
&\leq \sum_{i=1}^{\kappa_c t} \Big[ H(U_i) - H(U_i|W_i)  \Big]\\
\label{Eqn:BC:Gerhard-2}
&= \sum_{i=1}^{\kappa_c t} I(W_i;U_i)\ ,
\end{align}
where~\eqref{Eqn:BC:Gerhard-1} follows because $U_i \minuso W_i \minuso (W_1,W_2,\ldots,W_{i-1},W_{i+1},W_{i+2},\ldots,W_n,U_1,U_2,\ldots,U_{i-1})$ forms a Markov chain.  

Then we have
\begin{align}
\kappa_c I(\tilde{W};U) &\geq \frac{1}{t}\sum_{i=1}^{\kappa_c t} I(W_i;U_i) \\
\label{Eqn:BC:Gerhard-3}
&\geq \frac{1}{t}I(\mbf{W};\mbf{U})\\
\label{Eqn:BC-1}
&\geq \frac{1}{t}I(\mbf{X},\mbf{Y};\mbf{U})\\
&\geq \frac{1}{t}I(\mbf{X};\mbf{U}|\mbf{Y})\\
&= \frac{1}{t}\sum_{i=1}^{\kappa_s t} I(X_i;\mbf{U}|X_1^{i-1},\mbf{Y})\\
\label{Eqn:BC-2}
&= \frac{1}{t}\sum_{i=1}^{\kappa_s t} I(X_i;\mbf{U},X_1^{i-1},\mbf{Y}|Y_i)\\
\label{Eqn:BC:3}
&\geq \frac{1}{t}\sum_{i=1}^{\kappa_s t} I(X_i;\hat{X}_i|Y_i)\\
\label{Eqn:BC:4}
&\geq \frac{1}{t}\sum_{i=1}^{\kappa_s t} R_{X|Y}(d_{x,i})\\
\label{Eqn:BC:5}
&\geq \kappa_s R_{X|Y}(\delta_1)\\
\label{Eqn:BC:6}
&\geq \kappa_s R_{X|Y}(d_1+\epsilon)\ ,
\end{align}
where~\eqref{Eqn:BC-1} follows from the data-processing inequality, \eqref{Eqn:BC-2} follows because $(\mbf{X},\mbf{Y})$ is iid, \eqref{Eqn:BC:3} follows from the data-processing inequality and the fact that $\hat{X}_i$ is a function of $(\mbf{U},\mbf{Y})$, \eqref{Eqn:BC:4} follows from the definition of the conditional rate-distortion function where $d_{x,i} \triangleq \mathbb{E}\delta_1(X_i,\hat{X}_i)$, and \eqref{Eqn:BC:5} combines Jensen's inequality and the convexity of $R_{X|Y}(d_1)$ in $d_1$, and \eqref{Eqn:BC:6} follows because $R_{X|Y}(d_1+\epsilon)$ non-increasing in $d_1$. Similarly, it can be shown that
\begin{equation}
\kappa_c I(W;V) \geq \kappa_s R_{Y|X}(d_2 +\epsilon)\ .
\end{equation}
The theorem follows from the continuity of $R_{X|Y}(d_1)$ and $R_{Y|X}(d_2)$ on $\mathbb{R}_+$ and the fact that $\epsilon > 0$ is arbitrary.
\end{proof}

\medskip

\subsection{Achievability of Theorem~\ref{Thm:JSCC-Cut-Set}}

We now adapt an achievability result of Nayak, Tuncel and G$\ddot{\text{u}}$nd$\ddot{\text{u}}$z~\cite{Nayak-Apr-2010-A1} to give a sufficient condition for joint source-channel coding. When combined with Theorem~\ref{Thm:JSCC-Cut-Set}, this condition will give necessary and sufficient conditions for joint source-channel coding of jointly Gaussian random variables with squared-error distortion measures.

\medskip

\begin{lemma}[Cor. 1~\cite{Nayak-Apr-2010-A1}]\label{Lem:Nayak}
Let $\set{C}$ be a finite set. A distortion pair $(d_1,d_2) \in \mathbb{R}_+^2$ is admissible with bandwidth expansion $\kappa$ if the following conditions are satisfied:
\begin{enumerate}
\item[(i)] there exist random variables $W$ on $\set{W}$ and $C$ on $\set{C}$;
\item[(ii)] there exist functions $\pi_1: \set{C} \times \set{Y} \rightarrow \hat{\set{X}}$ and $\pi_2: \set{C} \times \set{X} \rightarrow \hat{\set{Y}}$ with
\begin{subequations}\label{Eqn:Nayak-Distortion}
\begin{align}
\mathbb{E}\big[\delta_1(X,\pi_1(C,Y))\big] &\leq d_1\\
\mathbb{E}\big[\delta_2(Y,\pi_2(C,X))\big] &\leq d_2\ ;
\end{align}
\end{subequations}
\item[(iii)] the following inequalities hold
\begin{subequations}
\begin{align}
I(X;C|Y) &\leq \kappa I(W;U)\\
I(Y;C|X) &\leq \kappa I(W;V)\ .
\end{align}
\end{subequations}
\end{enumerate}
\end{lemma}

\medskip

Lemma~\ref{Lem:Nayak} is the joint source-channel coding extension of the one-description upper bound given in Lemma~\ref{Lem:One-Description-Upper}. The lemma is actually a special case of a stronger
result~\cite[Thm. 1]{Nayak-Apr-2010-A1}; however, this weaker result will suffice for the following discussion. Note also the Markov constraints in~\cite[Cor.1]{Nayak-Apr-2010-A1} do not play a role here as the side-information is available to the transmitter.

The next two corollaries combine Theorem~\ref{Thm:JSCC-Cut-Set} and Lemma~\ref{Lem:Nayak} to give necessary and sufficient conditions for the following two special cases: $(i)$ the source $q_{XY}$ has zero-rate loss in the Wyner-Ziv problem, and $(ii)$ one source has to be reconstructed vanishing Hamming distortion.

\medskip

\begin{corollary}\label{Cor:JSCC:Zero-WZ-Rate-Loss}
If $q_{XY}$ has zero rate-loss in the Wyner-Ziv problem (i.e., $R_{X|Y}(d_1) = R_{X|Y}^{WZ}(d_1)$ and $R_{Y|X}(d_2) = R_{Y|X}^{WZ}(d_2)$), then $(d_1,d_2) \in \reals^2$ is achievable with bandwidth expansion $\kappa$ if and only if there exists a pmf $p_W$ on $\set{W}$ such that~\eqref{Eqn:ThmJSCC-Cut-Set} holds.
\end{corollary}

\medskip

As discussed before, the zero Wyner-Ziv rate-loss condition is very restrictive and few sources are known to satisfy it. However, an interesting example that does satisfy this condition is given next.

\medskip

\begin{example}
If $(X,Y)$ are jointly Gaussian random variables $\delta_1$ and $\delta_2$ are squared error distortion measures~\eqref{Eqn:Squared-Error} (see Example~\ref{Exa:RD-Bounds:Exa:1Gaussian}), then $(d_1,d_2) \in \reals^2$ is achievable with bandwidth expansion $\kappa$ if and only if there exists a pmf $p_W$ on $\set{W}$ such that~\eqref{Eqn:ThmJSCC-Cut-Set} holds. The conditional RD functions $R_{X|Y}(d_1)$ and $R_{Y|X}(d_2)$ are given in~\eqref{Eqn:CRD-Gaussian}.
\end{example}

\medskip

\begin{corollary}\label{Cor:JSCC:One-Lossless-Reconstruction}
If $\delta_1$ is a Hamming distortion measure, then $(0,d_2)$ is achievable with bandwidth expansion $\kappa$ if and only if there exists a pmf $p_W$ on $\set{W}$ such that
\begin{subequations}
\begin{align}
H(X|Y) &\leq \kappa I(W;U)\quad \text{and}\\
R_{Y|X}(d_2) &\leq \kappa I(W;V)\ .
\end{align}
\end{subequations}
\end{corollary}

\medskip

\begin{proof}[Proof of Corollary~\ref{Cor:JSCC:Zero-WZ-Rate-Loss}]
The necessary condition (``only if'') is given by Theorem~\ref{Thm:JSCC-Cut-Set}. The sufficient condition (``if'') is proved by constructing an auxiliary random variable $C$ that meets the conditions of Lemma~\ref{Lem:Nayak} with $I(X;C|Y) =R_{X|Y}(d_1)$ and $I(Y;C|X) = R_{Y|X}(d_2)$.
Recall that
\begin{align}
R_{X|Y}^{WZ}(d_1) &\triangleq \min_{p\in\set{P}_{X|Y}^{WZ}(d_1)} I(X;A|Y)\ , \text{ and}\\
R_{Y|X}^{WZ}(d_2) &\triangleq \min_{p\in\set{P}_{Y|X}^{WZ}(d_2)} I(Y;B|X)\ .
\end{align}
Let $p'$ and $p''$ be joint pmfs on $\set{A} \times \set{X} \times \set{Y}$ and $\set{B} \times \set{X} \times \set{Y}$ that achieve the aforementioned minima. Let $p$ be the joint pmf on $\set{A} \times \set{B} \times \set{X} \times \set{Y}$ defined by
\begin{equation}
p(a,b,x,y) \triangleq
\left\{
\begin{array}{ll}
\frac{p'(a,x,y) p''(b,x,y)}{q_{XY}(x,y)}, & \hbox{ if } q_{XY}(x,y) > 0\ , \\
0, & \hbox{ otherwise.}
\end{array}
\right.
\end{equation}
By construction, the $(A,X,Y)$ and $(B,X,Y)$ marginals of $p$ are $p'$ and $p''$, and $p$ satisfies the chain $A \minuso (X,Y) \minuso B$. Recall that $p'$ satisfies the chain $A \minuso X \minuso Y$, and $p''$ satisfies the chain $B \minuso Y \minuso X$. Combining these chains yields the long chain $A \minuso X \minuso Y \minuso B$.

Set $\set{C} \triangleq \set{A} \times \set{B}$ and $C = (A,B)$. Note that $C$ is a valid auxiliary random variable for Lemma~\ref{Lem:Nayak}. Moreover, we have
\begin{align}
I(X;C|Y) &= I(X;A,B|Y)\\
&= I(X;A|Y) + I(X;B|A,Y)\\
\label{Eqn:Proof:BC:Cor:1:1}
&= I(X;A|Y)\\
\label{Eqn:Proof:BC:Cor:1:2}
&= R_{X|Y}^{WZ}(d_1)\\
\label{Eqn:Proof:BC:Cor:1:3}
&= R_{X|Y}(d_1)\ ,
\end{align}
where~\eqref{Eqn:Proof:BC:Cor:1:1} follows because $A \minuso X \minuso Y \minuso B$ implies $X \minuso (A,Y) \minuso B$, \eqref{Eqn:Proof:BC:Cor:1:2} follows because $p'$ is an optimal test channel for the Wyner-Ziv RD function, and \eqref{Eqn:Proof:BC:Cor:1:3} follows by assumption. Similarly, we have $I(Y;C|X) = R_{Y|X}(d_2)$.
\end{proof}

\begin{proof}[Proof of Corollary~\ref{Cor:JSCC:One-Lossless-Reconstruction}]
The necessary condition (``only if'') is follows from Theorem~\ref{Thm:JSCC-Cut-Set} and $R_{X|Y}(0) = H(X|Y)$. The sufficient condition (``if'') is proved by constructing an auxiliary random variable $C$ that meets the conditions of Lemma~\ref{Lem:Nayak} as well as $I(X;C|Y) =R_{X|Y}(0)=H(X|Y)$ and $I(Y;C|X) = R_{Y|X}(d_2)$.

Recall the joint pmf of $(X,Y,\hat{X},\hat{Y})$ used to prove Corollary~\ref{Thm:CR-RD:Cor:OneLossless}. Choose $C = (\hat{X},\hat{Y})$ and note this choice of $C$ meets the conditions of Lemma~\ref{Lem:Nayak}. As before, we also have that $I(X;C|Y) = I(X;\hat{X},\hat{Y}|Y) = H(X|Y)$ and $I(Y;C|X) = I(Y;\hat{X},\hat{Y}|X) = R_{Y|X}(d_2)$.
\end{proof}

\subsection{Proof of Theorem~\ref{Thm:JSCC-CR}}

The sufficient condition is a special case of Lemma~\ref{Lem:Nayak}. We now give the necessary condition. If a distortion pair $(d_1,d_2) \in \reals^2$ is achievable with bandwidth expansion $\kappa = \kappa_c / \kappa_s$, then by definition there exists for every $\epsilon > 0$ a CR-JSC code $(f^{(t)},g_1^{(t)},g_2^{(t)},\phi_1^{(t)},\phi_2^{(t)})$ with
\begin{align}
\Delta_i^{(\kappa_s t)} &\leq d_i + \epsilon\ ,
\end{align}
as well as
\begin{align}
\label{Eqn:App-G:CR-Error-Prob-1}
\Pr\big[\phi_2^{(t)}(\mbf{V},\mbf{X}) \neq g_1^{(t)}(\mbf{U},\mbf{Y})\big] &\leq \epsilon_t  \text{ and}\\
\label{Eqn:App-G:CR-Error-Prob-2}
\Pr\big[\phi_1^{(t)}(\mbf{U},\mbf{Y}) \neq g_2^{(t)}(\mbf{V},\mbf{X})\big] &\leq \epsilon\ ,
\end{align}
As in the proof of Theorem~\ref{Thm:JSCC-Cut-Set}, let $\mbf{W} = f^{(t)}(\mbf{X},\mbf{Y})$, let $p_{W_i}$ denote the pmf for the $i^{\text{th}}$ symbol $W_i$, and define the time shared random variable $\tilde{W}$ on $\set{W}$ via
\begin{equation}
p_{\tilde{W}}(w) \triangleq \frac{1}{\kappa_c t} \sum_{i=1}^{\kappa_c t} p_{W_i}(w)\ .
\end{equation}
We will show that
\begin{subequations}
\begin{align}
\kappa_c I(\tilde{W};U) &\geq \kappa_s I(X;\hat{X},\hat{Y}|Y)\quad \text{ and}\\
\kappa_c I(\tilde{W};V) &\geq \kappa_s I(Y;\hat{X},\hat{Y}|X)
\end{align}
\end{subequations}
for some test channel $p_{\hat{X}\hat{Y}|XY} \in \set{P}_{\hat{X}\hat{Y}|XY}(d_1,d_2)$.

The next inequality, which will be useful later, follows from Fano's inequality~\cite{Cover-2006-B} and~\eqref{Eqn:App-G:CR-Error-Prob-2}:
\begin{align}\label{Eqn:App-G:vepsilon}
\varepsilon(\kappa_s, t, \epsilon)
&\geq \frac{1}{t}H\big(g_2^{(t)}(\mbf{V},\mbf{X})\big|\phi_1^{(t)}(\mbf{U},\mbf{Y})\big)\ ,
\end{align}
where
\begin{equation}
\varepsilon(\kappa_s, t, \epsilon) \triangleq \frac{1}{t}h(\epsilon) + \epsilon \kappa_s \log_2 |\hat{\set{X}}||\hat{\set{Y}}|\ .
\end{equation}

We first invoke the techniques used in the converse proof of Theorem~\ref{Thm:JSCC-Cut-Set}; specifically, we have
\begin{align*}
\notag
\kappa_c I(W;U) &\geq \frac{1}{t} \sum_{j=1}^{\kappa_c t} I(W_j;U_j)\\
\notag
&\geq \frac{1}{t} I(\mbf{W};\mbf{U})\\
\notag
&\geq \frac{1}{t} I(\mbf{X},\mbf{Y};\mbf{U})\\
&\geq \frac{1}{t} I(\mbf{X};\mbf{U}|\mbf{Y})\ .
\end{align*}

We now invoke the techniques used in the converse proof of Theorem~\ref{Thm:CR-RD}. Specifically, we have
\begin{align}
\kappa_c I(\tilde{W};U) &\geq \frac{1}{t} I(\mbf{X};\mbf{U}|\mbf{Y})\\
\label{Eqn:App-G:S-1}
&= \frac{1}{t} I\big(\mbf{X};\mbf{U},g_1^{(t)}(\mbf{U},\mbf{Y}),\phi_1^{(t)}(\mbf{U},\mbf{Y})\big|\mbf{Y}\big)\\
\notag
&= \frac{1}{t} \Big[ I\big(\mbf{X};\mbf{U},g_1^{(t)}(\mbf{U},\mbf{Y}),\phi_1^{(t)}(\mbf{U},\mbf{Y}),g_2^{(t)}(\mbf{V},\mbf{X})\big|\mbf{Y}\big)\\
&\qquad\qquad\qquad- I\big(\mbf{X};g_2^{(t)}(\mbf{V},\mbf{X})\big|\mbf{Y},\mbf{U},g_1^{(t)}(\mbf{U},\mbf{Y}),\phi_1^{(t)}(\mbf{U},\mbf{Y})\big)\Big]\\
\label{Eqn:App-G:S-2}
&\geq \frac{1}{t} I\big(\mbf{X};\mbf{U},g_1^{(t)}(\mbf{U},\mbf{Y}),\phi_1^{(t)}(\mbf{U},\mbf{Y}),g_2^{(t)}(\mbf{V},\mbf{X})\big|\mbf{Y}\big) - \varepsilon(\kappa_s, t,\epsilon)\\
\label{Eqn:App-G:S-3}
&\geq \frac{1}{t} I\big(\mbf{X};g_1^{(t)}(\mbf{U},\mbf{Y}),g_2^{(t)}(\mbf{V},\mbf{X})\big|\mbf{Y}\big) - \varepsilon(\kappa_s, t,\epsilon)\\
&= \frac{1}{t} \sum_{j = 1}^{\kappa_s t} I\big(X_j;g_1^{(t)}(\mbf{U},\mbf{Y}),g_2^{(t)}(\mbf{V},\mbf{X})\big|\mbf{Y},X_1^{j-1}\big) - \varepsilon(\kappa_s, t,\epsilon)\\
&= \frac{1}{t} \sum_{j = 1}^{\kappa_s t} I\big(X_j;g_1^{(t)}(\mbf{U},\mbf{Y}),g_2^{(t)}(\mbf{V},\mbf{X}),X_1^{j-1},Y_1^{j-1},Y_{j+1}^{\kappa_s t}\big|Y_j\big) - \varepsilon(\kappa_s, t,\epsilon)\\
\label{Eqn:App-G:S-4}
&\geq \frac{1}{t} \sum_{j = 1}^{\kappa_s t} I\big(X_j;g_1^{(t)}(\mbf{U},\mbf{Y}),g_2^{(t)}(\mbf{V},\mbf{X})\big|Y_j\big) - \varepsilon(\kappa_s, t,\epsilon)\ ,
\end{align}
where~\eqref{Eqn:App-G:S-1} follows because $\mbf{X} \minuso (\mbf{U},\mbf{Y}) \minuso (g_1^{(t)}(\mbf{U},\mbf{Y}),\phi_1^{(t)}(\mbf{U},\mbf{Y}))$ forms a Markov chain, and~\eqref{Eqn:App-G:S-2} follows from~\eqref{Eqn:App-G:vepsilon} and
\begin{align}
\varepsilon(\kappa_s, t,\epsilon) &\geq \frac{1}{t}H\big(g_2^{(t)}(\mbf{V},\mbf{X})\big|\phi_1^{(t)}(\mbf{U},\mbf{Y})\big)\\ &\geq \frac{1}{t}I\big(\mbf{X};g_2^{(t)}(\mbf{V},\mbf{X})\big|\mbf{Y},\mbf{U},\phi_1^{(t)}(\mbf{U},\mbf{Y}),g_1^{(t)}(\mbf{U},\mbf{Y})\big) \ .
\end{align}

A similar procedure yields
\begin{equation}\label{Eqn:App-G:S-5}
\kappa_c I(\tilde{W};U) \geq \frac{1}{t} \sum_{j = 1}^{\kappa_s t} I\big(Y_j;g_1^{(t)}(\mbf{U},\mbf{Y}),g_2^{(t)}(\mbf{V},\mbf{X})\big|X_j\big) - \varepsilon(\kappa_s, t,\epsilon)\ .
\end{equation}

For $i = 1,2,\ldots, \kappa_s t$, let $\hat{X}_i$ and $\hat{Y}_i$ denote the $i^{\text{th}}$ symbols of $g_1^{(t)}(\mbf{U},\mbf{Y})$ and $g_2^{(t)}(\mbf{V},\mbf{X})$, respectively. Let $p_{\hat{X}_j\hat{Y}_j|X_j,Y_j}(\hat{x}_j,\hat{y}_j|x_j,y_j)$ denote the conditional probability of $(\hat{X}_j,\hat{Y}_j)$ given $(X_j,Y_j)$, and define
\begin{equation}
p_{\hat{X}\hat{Y}|XY}(\hat{x},\hat{y}|x,y) \triangleq \frac{1}{\kappa_s t} \sum_{j=1}^{\kappa_s t} p_{\hat{X}_j\hat{Y}_j|XY}(\hat{x},\hat{y}|x,y)\ .
\end{equation}
The average distortion requirement on the code guarantees
\begin{align}
\sum_{x,y,\hat{x},\hat{y}} p_{\hat{X}\hat{Y}|XY}(\hat{x},\hat{y}|x,y)q_{XY}(x,y) \delta_1(x,\hat{x}) &\leq d_1 + \epsilon\quad \text{ and}\\
\sum_{x,y,\hat{x},\hat{y}} p_{\hat{X}\hat{Y}|XY}(\hat{x},\hat{y}|x,y)q_{XY}(x,y) \delta_2(y,\hat{y}) &\leq d_2 + \epsilon\ .
\end{align}

We further have
\begin{align}
\label{Eqn:Thm:JSCC:Rate-3}
\frac{1}{t} \sum_{j=1}^{\kappa_s t} I\big(X_j;\hat{X}_j,\hat{Y}_j\big|Y_j\big) &\geq \kappa_s I(X;\hat{X},\hat{Y}|Y)\ , \text{ and}\\
\label{Eqn:Thm:JSCC:Rate-4}
\frac{1}{t} \sum_{j=1}^{\kappa_s t} I\big(Y_j;\hat{X}_j,\hat{Y}_j\big|X_j\big) &\geq \kappa_s I(Y;\hat{X},\hat{Y}|X)\ ,
\end{align}
where we have used Jensen's inequality together with the convexity of $I(X;\hat{X},\hat{Y}|Y)$ and $I(Y;\hat{X},\hat{Y}|X)$ in $p_{\hat{X}\hat{Y}|XY}$. Thus, we have shown that there exists a condition pmf $p_{\hat{X}\hat{Y}|XY}(\hat{x},\hat{y}|x,y)$ and a pmf $p_{\tilde{W}}$ such that
\begin{align}
\kappa_c I(\tilde{W};U) &\geq \kappa_s I(X;\hat{X},\hat{Y}|Y) - \varepsilon(\kappa_s, t,\epsilon)\quad \text{ and}\\
\kappa_c I(\tilde{W};V) &\geq \kappa_s I(Y;\hat{X},\hat{Y}|X) - \varepsilon(\kappa_s, t,\epsilon)\ .
\end{align}

\subsection{Proof of Theorem~\ref{Thm:JSCC-Small-Distortions}} 

The necessary condition follows from Theorem~\ref{Thm:JSCC-Cut-Set}. We now show that this necessary condition is also sufficient for small distortions. From Theorem~\ref{Thm:JSCC-CR}, a sufficient condition for $(d_1,d_2) \in \reals^2$ to be achievable is that there exists a pmf $p_W$ on $\set{W}$ and $p_{\hat{X}\hat{Y}|XY} \in \set{P}_{\hat{X}\hat{Y}|XY}(d_1,d_2)$ such that~\eqref{Eqn:Thm:JSCC-CR-1} holds. Choose $p_{\hat{X}\hat{Y}|XY} \in \set{P}_{\hat{X}\hat{Y}|XY}(d_1,d_2)$ to achieve the minimum in the definition of the joint RD function $R_{XY}(d_1,d_2)$. In a similar manner to the proof of Corollary~\ref{Cor:Thm:CR-RD}, we have that
\begin{subequations}
\begin{align}
I(X;\hat{X},\hat{Y}|Y) &\leq R_{XY}(d_1,d_2) - R_Y(d_2)\quad \text{and} \\
I(Y;\hat{X},\hat{Y}|X) &\leq R_{XY}(d_1,d_2) - R_X(d_1)\ .
\end{align}
\end{subequations}
From Gray~\cite[Thm. 3.2]{Gray-Jul-1973-A}, there exists a strictly positive surface $\set{D}$ in $\reals^2$ such that $R_{X|Y}(d_1) = R_{XY}(d_1,d_2) - R_Y(d_2)$ and $R_{Y|X}(d_2) = R_{XY}(d_1,d_2) - R_X(d_1)$ whenever $(d_1,d_2)$ lies on or below $\set{D}$. For these small distortions, we have that $I(X;\hat{X},\hat{Y}|Y) = R_{X|Y}(d_1)$ and $I(Y;\hat{X},\hat{Y}|X) = R_{Y|X}(d_2)$.


\section{Conclusion}\label{Sec:6}
The downlink broadcast channel of the two-way relay network was studied in the source coding and joint source-channel coding settings. Single-letter necessary and sufficient conditions for reliable communication were given for the following special cases: common-reconstructions (Theorems~\ref{Thm:CR-RD} and~\ref{Thm:JSCC-CR}), small distortions (Theorems~\ref{Thm:Small-Distortions} and~\ref{Thm:JSCC-Small-Distortions}), conditionally independent sources (Corollary~\ref{Thm:RD-Bounds:Cor:Cond-Ind}), deterministic distortion measures (Corollary~\ref{Thm:RD-Bounds:Cor:2Deterministic}), and sources with zero rate-loss for the Wyner-Ziv problem~\cite{Zamir-Nov-1996-A}. Additionally, the notion of small distortions was explicitly characterised for the doubly symmetric binary source with Hamming distortion measures in Theorem~\ref{Thm:DSBS-CR}. 
Each of the aforementioned results followed, in part, from the necessary conditions presented in Theorems~\ref{Thm:JSCC-Cut-Set} and~\ref{Thm:RD-Bounds}. It remains to be verified that these necessary conditions are, or are not, sufficient. 

More generally, the source coding problem is a special case of the Wyner-Ziv problem with two receivers~\cite{Heegard-Nov-1985-A,Timo-Jun-2010-C}, and the joint source-channel coding problem is a special case of the Wyner-Ziv coding over broadcast channels problem~\cite{Nayak-Apr-2010-A1}. It would be interesting to see if the small distortion results in this paper carry over to these problems. 


\section*{Acknowledgements}
The authors are grateful to Gottfried Lechner, Badri N. Vellambi, Terence Chan and Tobias Oechtering for many interesting discussions on the contents of this paper.

\appendices


\section{An Improved Lower Bound for $R(d_1,d_2)$}\label{App:NewLowerBound}

In this section, we present an alternative to the cut-set lower bound $R_L(d_1,d_2)$ given in Theorem~\ref{Thm:RD-Bounds} (see the end of Section~\ref{Sec:4A}). For this purpose, let $\set{A}$, $\set{B}$ and $\set{C}$ be finite alphabets of cardinality
\begin{subequations}
\begin{align}
|\set{C}| &\leq |\set{X}|\ |\set{Y}| + 5\ ,\\
|\set{A}| &\leq |\set{X}|\ |\set{Y}|\ |\set{C}| + 2\ \text{ and}\\
|\set{B}| &\leq |\set{X}|\ |\set{Y}|\ |\set{C}| + 2\ .
\end{align}
\end{subequations}
The new lower bound will be obtained by minimizing a certain function over the following set of joint pmfs. Let $\set{P}_{L}^*(d_1,d_2)$ denote the set of pmfs $p$ on $\set{A}$ $\times$ $\set{B}$ $\times$ $\set{C}$ $\times$ $\set{X}$ $\times$ $\set{Y}$ where
\begin{enumerate}
\item[(i)] $A \minuso (X,Y,C) \minuso B$ forms a Markov chain, i.e.,
\begin{equation}
I(A;B|X,Y,C) = 0\ ,
\end{equation}
\item[(ii)] $(A,B)$ is independent of $(X,Y)$, i.e.,
\begin{equation}
I(X,Y;A,B) = 0\ ,
\end{equation}
\item[(iii)] there exist functions
\begin{subequations}
\begin{align}
\pi_1 :& \set{A} \times \set{C} \times \set{Y} \rightarrow \Hat{\set{X}} \text{ and}\\
\pi_2 :& \set{B} \times \set{C} \times \set{X} \rightarrow \Hat{\set{Y}}
\end{align}
\end{subequations}
such that
\begin{subequations}
\begin{align}
\mathbb{E}_p\Big[\delta_1\big(X,\pi_1(A,C,Y)\big)\Big] &\leq d_1, \text{ and}\\
\mathbb{E}_p\Big[\delta_2\big(Y,\pi_2(B,C,X)\big)\Big] &\leq d_2\ .
\end{align}
\end{subequations}
\end{enumerate}
Define
\begin{align}
R_{L}^*(d_1,d_2) &\triangleq \min_{p \in \set{P}_L^*(d_1,d_2)} \Bigg[ I(X,Y;C) + \max \Big\{I(X;A|C,Y),\ I(Y;B|C,X) \Big\} \Bigg]\ .
\end{align}
The next theorem gives a lower bound for $R(d_1,d_2)$.

\medskip

\begin{theorem}\label{Thm:RD*-Lower-Bound}
For $(d_1,d_2) \in \reals^2$ we have that
\begin{equation}
R(d_1,d_2) \geq R_L^*(d_1,d_2) \geq R_{L}(d_1,d_2)\ .
\end{equation}
\end{theorem}

\medskip

\subsection{Proof: $R(d_1,d_2) \geq R^*_{L}(d_1,d_2)$}

If $r$ is $(d_1,d_2)$-admissible, then there exists a monotonically decreasing sequence $\{\epsilon_i\}$ with limit zero; a monotonically increasing sequence $\{n_i\}$; and a sequence of RD codes $\{(f^{(n_i)},g_1^{(n_i)},g_2^{(n_i)})\}$ such that $\kappa^{(n_i)} \leq r + \epsilon_i$, $\Delta_1^{(n_i)} \leq d_1 + \epsilon_i$ and $\Delta_2^{(n_i)} \leq d_2 + \epsilon_i$.
Then we have
\begin{align}
\label{RD-Converse-8}
r +& \epsilon \geq \frac{1}{n_i}\log_2\big|\set{M}^{(n_i)}\big|\\
\label{RD-Converse-10}
&= \frac{1}{n_i} H(M)\\
\label{RD-Converse-11}
&\geq \frac{1}{n_i} I(\mbf{X},\mbf{Y};M)\\
\label{RD-Converse-12}
&= \frac{1}{n_i}\Big[I(\mbf{X};M|\mbf{Y}) + I(\mbf{Y};M)\Big]\\
\label{RD-Converse-13}
&= \frac{1}{n_i}\sum_{j=1}^{n_i}\Big[ I(X_j;M|X_1^{j-1},\mbf{Y}) + I(Y_j;M|Y_1^{j-1})\Big]\\
\label{RD-Converse-14}
&= \frac{1}{n_i}\sum_{j=1}^{n_i}\Big[ I(X_j;M,X_1^{j-1},Y_1^{j-1},Y_{j+1}^{n_i}|Y_j) + I(Y_j;M,Y_1^{j-1})\Big]\\
\label{RD-Converse-15}
&\geq \frac{1}{n_i}\sum_{j=1}^{n_i}\Big[ I(X_j;M,Y_1^{j-1},Y_{j+1}^{n_i}|Y_j) + I(Y_j;M)\Big]\\
\label{RD-Converse-16}
&= \frac{1}{n_i}\sum_{j=1}^{n_i}\Big[ I(X_j;M|Y_j) + I(X_j;Y_1^{j-1},Y_{j+1}^{n_i}|M,Y_j) + I(Y_j;M)\Big]\\
\label{RD-Converse-17}
&= \frac{1}{n_i}\sum_{j=1}^{n_i}\Big[ I(X_j,Y_j;M) + I(X_j;Y_1^{j-1},Y_{j+1}^{n_i}|M,Y_j) \Big]\ ,
\end{align}
where~\eqref{RD-Converse-8} follows from the definition of a $(d_1,d_2)$-admissible rate, \eqref{RD-Converse-10} through \eqref{RD-Converse-13} follow from standard identities, \eqref{RD-Converse-14} follows because $(\mbf{X},\mbf{Y})$ is i.i.d., \eqref{RD-Converse-15} through \eqref{RD-Converse-17} follows from standard identities. In a similar manner, it can also be shown that
\begin{equation}\label{RD-Converse-18}
r + \epsilon \geq \frac{1}{n_i}\sum_{j=1}^{n_i}\Big[ I(X_j,Y_j;M) + I(Y_j;X_1^{j-1},X_{j+1}^{n_i}|M,X_j) \Big]\ .
\end{equation}

For $j = 1,2,\ldots,{n_i}$, define $\set{A}_j \triangleq \set{Y}^{{n_i}-1}$, $\set{B}_j \triangleq \set{X}^{{n_i}-1}$, and $\set{C}_j \triangleq \set{M}^{({n_i})}$. We consider $\{\set{C}_j\}$, $j = 1,2,\ldots,n_i$, to be a class of disjoint sets. Similarly, we consider $\{\set{A}_j\}$ and $\{\set{B}_j\}$ to be disjoint sets. Now define
\begin{subequations}
\begin{align}
A_j &\triangleq (Y_1^{j-1},Y_{j+1}^{n_i})\ ,\\
B_j &\triangleq (X_1^{j-1},X_{j+1}^{n_i})\ \text{ and}\\
C_j &\triangleq M\ ,
\end{align}
\end{subequations}
Let $p^*_j$ denote the resultant joint pmf on $\set{A}_j$ $\times$ $\set{B}_j$ $\times$ $\set{C}_j$ $\times$ $\set{X}$ $\times$ $\set{Y}$ that characterises the random variables $A_j$, $B_j$, $C_j$, $X_j$ and $Y_j$. By construction, we have
\begin{enumerate}
\item $(X_j,Y_j)$ is independent of $(A_j,B_j)$, i.e.
\begin{equation}
I_{p^*_j}(X_j,Y_j;A_j,B_j) = I(X_j,Y_j;Y_1^{j-1},Y_{j+1}^n,X_1^{j-1},X_{j+1}^n) = 0\ ,
\end{equation}
\item there exists a function $\pi_{x,j}: \set{A}_j \times \set{C}_j \times \set{Y} \rightarrow \Hat{\set{X}}$ such that $\Hat{X}_j = \pi_{x,j}(A_j,C_j,Y_j)$,
\item there exists a function $\pi_{y,j}: \set{B}_j \times \set{C}_j \times \set{X} \rightarrow \Hat{\set{Y}}$ such that $\Hat{Y}_j = \pi_{y,j}(B_j,C_j,X_j)$.
\end{enumerate}
Now define $\set{A} \triangleq \cup_j \set{A}_j$, $\set{B} \triangleq \cup_j \set{B}_j$ and $\set{C} \triangleq \cup_j \set{C}_j$, and the ``time-shared'' pmf
\begin{equation*}
p^*(x,y,a,b,c) \triangleq
\left\{
  \begin{array}{ll}
    \frac{1}{n_i} p_j^*(a,b,c,x,y)\ , & \hbox{ if } a \in \set{A}_j,\\
 &\quad b \in \set{B}_j,\ c \in \set{C}_j,\\
    0\ , & \hbox{ otherwise.}
  \end{array}
\right.
\end{equation*}
on $\set{A} \times \set{B} \times \set{C} \times \set{X} \times \set{Y}$. Using this definition, it can be verified that
\begin{subequations}
\begin{align}
\label{Eqn:App-B:TS1}
I_{p^*}(X,Y;C) &= \frac{1}{n_i}\sum_{j=1}^{n_i} I_{p_j^*}(X_j,Y_j;C_j)\\
\label{Eqn:App-B:TS2}
I_{p^*}(X;A|C,Y) &= \frac{1}{n_i}\sum_{j=1}^{n_i} I_{p_j^*}(X_j;A_j|C_j,Y_j)\\
\label{Eqn:App-B:TS3}
I_{p^*}(X;A|C,Y) &= \frac{1}{n_i}\sum_{j=1}^{n_i} I_{p_j^*}(X_j;A_j|C_j,Y_j)\\
I_{p^*}(X,Y;A,B) &= \frac{1}{n_i} \sum_{j=1}^{n_i} I(X_j,Y_j;A_j,B_j) = 0\ .
\end{align}
\end{subequations}
Furthermore, by definition, we have
\begin{align}
d_1 + \epsilon_i &\geq \Delta_1^{(n_i)}\\
&= \frac{1}{n_i}\sum_{j=1}^{n_i} \mathbb{E}_{p_j^*}\Big[\delta_1\big(X_j,\pi_{1,j}(A_j,C_j,Y_j)\big)\Big]\\
&= \mathbb{E}_{p^*}\big[\delta_1(X,\pi_1(A,C,Y)\big]\ ,
\end{align}
where the last expectation is taken with respect to $p^*$, and $\pi_1:\set{A} \times \set{C} \times \set{Y} \rightarrow \hat{\set{X}}$ is defined by
\begin{equation}
\pi_1(a,c,y) \triangleq \left\{
                          \begin{array}{ll}
                            \pi_{1,j}(a,c,y) & \hbox{ if } a \in \set{A}_j\ , \ c \in \set{C}_j \\
                            \hat{x}^*, & \hbox{ otherwise,}
                          \end{array}
                        \right.
\end{equation}
where $\hat{x}^* \in \hat{\set{X}}$ is arbitrary. Similarly,
\begin{equation}
d_2 + \epsilon_i \geq \mathbb{E}\big[\delta_2(Y,\pi_2(B,C,X))\big]\ .
\end{equation}

At this point we have that
\begin{align}
r + \epsilon_i \geq \inf \Big[& I_{p^*}(X,Y;C) + \max\big\{I_{p^*}(X;A|C,Y) , I_{p^*}(Y;B|C,X)\big\}\Big]\ ,
\end{align}
where the infimum is taken over all $p^*$ satisfying $I_{p^*}(X,Y;A,B) = 0$ as well as
\begin{subequations}
\begin{align}
d_1 + \epsilon_i &\geq \mathbb{E}_{p^*}\big[\delta_1(X,\pi_1(A,C,Y))\big]\ \text{ and}\\
d_2 + \epsilon_i &\geq \mathbb{E}_{p^*}\big[\delta_2(Y,\pi_2(B,C,X))\big]\ .
\end{align}
\end{subequations}
Note the this infimum is not altered if we impose the Markov chain $A \minuso (X,Y,C) \minuso B$. Finally, we apply the support lemma~\cite{Csiszar-1981-B} to bound the cardinality of $\set{C}$ by $|\set{X}|\ |\set{Y}| + 5$, and $\set{A}$ and $\set{B}$ by $|\set{X}|\ |\set{Y}|\ |\set{C}| + 2$. ($|\set{A}|$ and $|\set{B}|$ can be bounded simultaneously since  $A \minuso (X,Y,C) \minuso B$ forms a Markov chain.)

\medskip

\subsection{Proof: $R^*(d_1,d_2) \geq R_{L}(d_1,d_2)$}

Enlarge the set $\set{P}_{L}^*(d_1,d_2)$ by removing the constraints $I(X,Y;A,B) = 0$ and $A \minuso (X,Y,C) \minuso B$ $[p]$. Denote this new set by $\set{P}_{L}^\dag(d_1,d_2)$. Then,
\begin{align}
\notag
\min_{p \in \set{P}_{L}^*(d_1,d_2)}&  \Bigg[ I(X,Y;C) + \max \Big\{I(X;A|C,Y),\ I(Y;B|C,X) \Big\} \Bigg]\\
\label{Eqn:RD*-Converse-B-1}
&\geq \min_{p \in \set{P}_{L}^\dag(d_1,d_2)}  \Bigg[ I(X,Y;C) + \max \Big\{I(X;A|C,Y),\ I(Y;B|C,X) \Big\} \Bigg]\\
&= \min_{p \in \set{P}_{L}^\dag(d_1,d_2)}  \max \Big\{I(X,Y;C) + I(X;A|C,Y),\ I(X,Y;C) + I(Y;B|C,X) \Big\} \Bigg]\\
\label{Eqn:RD*-Converse-B-2}
&\geq \min_{p \in \set{P}_{L}^\dag(d_1,d_2)}  \max \Big\{I(X;A,C|Y),\ I(Y;B,C|X) \Big\} \Bigg]\\
\label{Eqn:RD*-Converse-B-3}
&= \min_{p \in \set{P}_{L}^\dag(d_1,d_2)}  \max \Big\{I(X;A,C,Y|Y),\ I(Y;B,C,X|X) \Big\} \Bigg]\\
\label{Eqn:RD*-Converse-B-4}
&\geq \min_{p \in \set{P}_{L}^\dag(d_1,d_2)}  \max \Big\{I(X;\pi_x(A,C,Y)|Y),\ I(Y;\pi_y(B,C,X)|X) \Big\} \Bigg]\\
\label{Eqn:RD*-Converse-B-5}
&\geq \max \Big\{R_{X|Y}(d_1),\ R_{Y|X}(d_2) \Big\} \\
\label{Eqn:RD*-Converse-B-6}
&\equiv R_{L}(d_1,d_2)\ .
\end{align}
where~\eqref{Eqn:RD*-Converse-B-1} follows because $\set{P}_{L}^*(d_1,d_2) \subseteq \set{P}_{L}^\dag(d_1,d_2)$, \eqref{Eqn:RD*-Converse-B-2} and~\eqref{Eqn:RD*-Converse-B-3} follow from the chain rule for mutual information, ~\eqref{Eqn:RD*-Converse-B-4} follows the data processing inequality, \eqref{Eqn:RD*-Converse-B-5} follows from the definition of the conditional rate-distortion function, and \eqref{Eqn:RD*-Converse-B-6} follows from the definition of $R_{L}(d_1,d_2)$. 


\section{Convexity of Conditional Mutual Information}\label{App:Lemma-Convexity}

\begin{proof}
Suppose $(A,B)$ is defined by a (fixed) joint pmf $p_{AB}$. Let $p^{(1)}_{C|AB}$ and $p^{(2)}_{C|AB}$ be two conditional pmfs for $C$ given $(A,B)$. For $i = 1,2$, let
\begin{equation}
p^{(i)}_{ABC} \triangleq p^{(i)}_{C|AB}(c|a,b)\ p_{AB}(a,b)\ , \quad (a,b,c) \in \set{A} \times \set{B} \times \set{C}
\end{equation}
denote the resulting joint pmfs. We identify the marginals of these pmfs with subscripts in the usual way; for example,
\begin{equation}
p^{(i)}_{AC}(a,c) \triangleq \sum_{b\in\set{B}} p^{(i)}_{ABC}(a,b,c)\ , \quad (a,c) \in \set{A} \times \set{C}\ , i = 1,2 \ .
\end{equation}

Choose $\alpha_1$ and $\alpha_2$ such that $0\leq \alpha_1,\alpha_2\leq 1$ and $\alpha_1 + \alpha_2 = 1$. Let
\begin{equation}
p^*_{C|AB}(c|a,b) \triangleq \alpha_1 p^{(1)}_{C|AB}(c|a,b) + \alpha_2 p^{(2)}_{C|AB}(c|a,b) \ .
\end{equation}
As before, let $p^*_{ABC}$ denote the resultant joint pmf for $(A,B,C)$ when $p^*_{C|AB}$ is combined with $p_{AB}$.

We wish to evaluate the conditional mutual information $I(A;C|B)$ with respect the three conditional probabilities\footnote{Note that $p^{(i)}_{AB}(a,b) = P^*_{AB}(a,b) = p_{AB}(a,b)$}: $p^{(1)}_{C|AB}$, $p^{(2)}_{C|AB}$ and $p^{*}_{C|AB}$. In particular, the lemma will be proved if it can be shown that
\begin{equation}
\sum_{i=1}^2 \alpha_i I(A;C|B)[p^{(i)}_{C|AB}] \geq I(A;C|B)[p^{*}_{C|AB}]\ ,
\end{equation}
where $I(A;C|B)[p'_{C|AB}]$ should be understood as the conditional mutual information $I(A;C|B)$ when the joint probability of $(A,B,C)$ is defined by $p_{AB}$ and $p'_{C|AB}$. For this purpose, we write $I(A;C|B)$ explicitly as a function of $p'_{C|AB}$:
\begin{equation}
I(A;C|B) = \sum_{a,b,c} p_B(b) p_{A|B}(a|b) p'_{C|AB}(c|a,b) \log
\frac{p'_{C|AB}(c|a,b)}{p'_{C|B}(c|b)}\ ,
\end{equation}
where the conditional probability $p'_{C|B}$ is a function of the other arguments
\begin{equation}
p'_{C|B}(c|b) = \sum_a p'_{A|B}(a|b)p^{(i)}_{C|AB}(c|a,b)\ .
\end{equation}

Then we have
\begin{align}
\sum_{i=1}^2 \alpha_i I(A;C|B)\ [p^{(i)}_{C|AB}] &= \sum_{i=1}^2
\sum_{a,b,c} \alpha_i p_B(b) p_{A|B}(a|b) p^{(i)}_{C|AB}(c|a,b) \log
\frac{p^{(i)}_{C|AB}(c|a,b)}{p^{(i)}_{C|B}(c|b)} \\
&= \sum_b p_B(b) \sum_{i=1}^2 \alpha_i \sum_{a,c} p_{A|B}(a|b) p^{(i)}_{C|AB}(c|a,b) \log
\frac{p^{(i)}_{C|AB}(c|a,b)}{p^{(i)}_{C|B}(c|b)}\\
&\geq \sum_b p_B(b) \sum_{a,c} p_{A|B}(a|b) p^{*}_{C|AB}(c|a,b) \log
\frac{p^{*}_{C|AB}(c|a,b)}{p^{*}_{C|B}(c|b)} \\
&= I(A;C|B)\ [p^*_{C|AB}]\ ,
\end{align}
where the inequality follows from the convexity of mutual information in the channel for a fixed input distribution~\cite{Cover-2006-B}.
\end{proof} 


\end{document}